\documentclass[preprint]{revtex4}
\usepackage[T1]{fontenc}
\usepackage{graphicx}
\usepackage{rotating}
\usepackage{bm}        % \boldsymbol (bold math)
\usepackage{amssymb}   % \mathbb     (blackboard bold)
\usepackage{bm}
\usepackage{float}
\usepackage{tikz}

\def\jpca#1#2#3{{\it J.~Phys.~Chem.~{\rm A}}~{\bf #1},\ #2\ (#3)}
\def\jpcc#1#2#3{{\it J.~Phys.~Chem.~{\rm C}}~{\bf #1},\ #2\ (#3)}
\def\jcp#1#2#3{{\it J.~Chem.~Phys.}~{\bf #1},\ #2\ (#3)}

\def\prl#1#2#3{{\it Phys.~Rev.~Lett.}~{\bf #1},\ #2\ (#3)}

\def\mp#1#2#3{{\it Mol. Phys.}~{\bf #1},\ #2\ (#3)}
\def\jpb#1#2#3{{\it J. Phys. B: At. Mol. Opt. Phys.} {\bf #1},\ #2\ (#3)}

\def\pccp#1#2#3{{\it Phys. Chem. Chem. Phys.}~{\bf #1},\ #2\ (#3)}
\def\irpc#1#2#3{{\it Int. Rev. Phys. Chem.}~{\bf #1},\ #2\ (#3)}
\def\jctc#1#2#3{{\it J. Chem. Theor. Comp.}~{\bf #1},\ #2\ (#3)}
\def\ijqc#1#2#3{{\it Int. J. Quant. Chem.}~{\bf #1},\ #2\ (#3)}

\def\k1{k_1}
\def\k2{k_2}
\def\q1{q_1}
\def\q2{q_2}

\def\({\left (}
\def\){\right )}
\def\[{\left [}
\def\]{\right ]}

\newcommand{\beq}{\begin{equation}}
\newcommand{\eeq}{\end{equation}}

\begin{document}
\date{\today}
\flushbottom \draft
\title{Interpolation and extrapolation of global potential energy surfaces for polyatomic systems 
by Gaussian processes with composite kernels}
\author{J. Dai and R. V. Krems}
\affiliation{
Department of Chemistry, University of British Columbia, Vancouver, B.C. V6T 1Z1, Canada
}
\begin{abstract}
Gaussian process regression has recently emerged as a powerful, system-agnostic tool for building global potential energy surfaces (PES) of polyatomic molecules. 
While the accuracy of GP models of PES increases with the number of potential energy points, so does the numerical difficulty of training and evaluating GP models. 
Here, we demonstrate an approach to improve the accuracy of global PES without increasing the number of energy points. 
The present work reports four important results. First, we show that the selection of the best kernel function for GP models of PES can be automated using the Bayesian information criterion as a model selection metric. 
Second, we demonstrate that GP models of PES trained by a small number of energy points can be significantly improved by iteratively increasing the complexity of GP kernels. 
The composite kernels thus obtained maximize the accuracy of GP models for a given distribution of potential energy points. Third, we show that the accuracy of the GP models of PES with composite kernels
can be further improved by varying the training point distributions. Fourth, we show that GP models with composite kernels can be used for physical extrapolation of PES. We illustrate the approach by constructing the six-dimensional PES for H$_3$O$^+$. 
For the interpolation problem, we show that this algorithm produces a global six-dimensional PES for H$_3$O$^+$ in
the energy range between zero and $21,000$ cm$^{-1}$ with the  root mean square error $65.8$ cm$^{-1}$ using only 500 randomly selected {\it ab initio} points as input.  To illustrate extrapolation, we produce the PES at high energies using the energy points at low energies as input. We show that one can obtain an accurate global fit of the PES extending to $21,000$ cm$^{-1}$ based on 1500 potential energy points at energies below $10,000$ cm$^{-1}$. 

\end{abstract}

\maketitle

\clearpage
\newpage

\section{Introduction}

Quantum properties of polyatomic molecules are computed by solving the nuclear Schr\"{o}dinger equation that is parametrized by potential energy surfaces (PES). These PES are generally obtained by fitting the results of the electronic structure calculations at different points of the nuclear configuration space. Many different techniques have been developed for fitting multi-dimensional PES (for representative examples, see Refs. \cite{fitting-PES,general-fitting-1,general-fitting-2,general-fitting-4,general-fitting-5,BML}). A major thrust of recent research has been to develop approaches for fitting PES of polyatomic systems by machine learning (ML) models \cite{ML-for-PES,NNs-for-PES, NNs-for-PESa,NNs-for-PES-1a,NNs-for-PES-1b,NNs-for-PES-1c,NNs-for-PES-2,NNs-for-PES-3,NNs-for-PES-4,NNs-for-PES-5,NNs-for-PES-6,gp-1,gp-2,gp-3,jie-jpb,gp-for-PES-2,gp-for-PES-3,gp-for-PES-4,gp-for-PES-5,gp-for-PES-6,gp-for-PES-7,gp-for-PES-8,gp-for-PES-9,gp-for-PES-10}. These models can be classified into approaches using artificial neural networks (NNs) \cite{ML-for-PES,NNs-for-PES, NNs-for-PESa,NNs-for-PES-1a,NNs-for-PES-1b,NNs-for-PES-1c,NNs-for-PES-2,NNs-for-PES-3,NNs-for-PES-4,NNs-for-PES-5,NNs-for-PES-6} and methods based on kernels \cite{general-fitting-2, BML,rabitz-1,rabitz-2,rabitz-3}, such as kernel ridge regression or Gaussian process regression \cite{gp-1,gp-2,gp-3,jie-jpb,gp-for-PES-2,gp-for-PES-3,gp-for-PES-4,gp-for-PES-5,gp-for-PES-6,gp-for-PES-7,gp-for-PES-8,gp-for-PES-9,gp-for-PES-10}. ML models have become popular because (i) they provide flexible representations of PES that can potentially be applied to molecular systems of any complexity; (ii) they are easy to construct given the abundance of relevant packages readily available.

The NN and kernel-based models of PES are generally designed to make accurate predictions of the potential energy {\it within} the range of the given potential energy points. NNs achieve this by an  analytical fit of the given points, while kernel-based methods interpolate the given points.  
However, the accuracy and numerical complexity of quantum chemistry calculations depends on the geometry of a given polyatomic system. For example, it is much more difficult to compute the potential energy at large intermolecular separations or where multiple electronic states exhibit degeneracies. 
At the same time, for complex molecular systems, the specific features of the PES in different parts of the configuration space are not {\it a priori} known. 
Therefore, for complex molecules, it is not generally known how to ensure that all important features of the PES fall {within} the range of the electronic structure calculations. An illustrative example of this challenge is the PES for the collision system of two alkali metal molecules \cite{nak}, of importance to ultracold chemistry experiments \cite{my-book}. For such and more complex molecular systems,  it would be desirable to develop ML approaches that could construct PES by extrapolation of a small number of {\it ab initio} points randomly placed in the configuration space. Such models of PES could be used to explore the entire landscape of the PES and inform further {\it ab initio} calculations to generate flexible and highly accurate PES for complex systems. At present, there is no general procedure for building global PES by extrapolation.  For example, if the PES is known only at low energies, it is generally considered unfeasible to build a global PES that is accurate at high energies in the entire configuration space. One of the goals of the present work is to demonstrate the feasibility of building ML models of PES by extrapolation.

%If an asymptotic behaviour of a PES is known, ML models can be forced to follow a specific analytical behaviour, either by restricting the kernels \cite{general-fitting-2} or by merging the model smoothly with an analytical fit \cite{nak}. However, if the asymptotic behaviour is not known, there is no general procedure for building global PES by extrapolation. For example, if the PES is known only at low energies, it is generally considered unfeasible to build a global PES that is accurate at high energies in the entire configuration space.  This makes the construction of PES for complex molecules difficult because one has to consider carefully how to place the electronic structure calculations in the configuration space.  For complex molecules, the specific features of the PES in different parts of the configuration space are not {\it a priori} known. 

The focus of the present work is on GP regression. While NNs provide parametric models of PES, GP models are nonparamteric. A nonparametric model can be viewed as one with an infinite number of parameters, whose distributions are adjusted in response to given potential energy points \cite{BML} in order to yield an accurate representation of a PES, on average. Using GPs to construct PES for polyatomic systems has several advantages over other methods \cite{BML}. First, it has been shown that accurate GP models can be obtained with a much smaller number of potential energy points than required for any other fitting method \cite{jie-jpb,gp-for-PES-4}.  Second, the construction of the models does not require any knowledge of the functional form of the PES and is completely automated. The same code can be applied to the construction of PES for different molecules of different complexity. 
This makes GP models ideal for applications aiming to invert the scattering problem \cite{rodrigo-bo} or where the PES is iteratively improved by adding more {\it ab initio} calculations to relevant parts of the configuration space. 
Finally, there is no effort required to control overfitting in GP models of PES, at least, for systems with less than 15 degrees of freedom that have been fitted with GP models in the literature so far \cite{jie-jpb,gp-for-PES-2,gp-for-PES-3,gp-for-PES-4,gp-for-PES-5,gp-for-PES-6,gp-for-PES-7,gp-for-PES-8,gp-for-PES-9,gp-for-PES-10}.

The present work reports several significant results. We first consider the interpolation problem, as in previous work \cite{gp-1,gp-2,gp-3,jie-jpb,gp-for-PES-2,gp-for-PES-3,gp-for-PES-4,gp-for-PES-5,gp-for-PES-6,gp-for-PES-7,gp-for-PES-8,gp-for-PES-9,gp-for-PES-10}, and show that the accuracy of the resulting GP models of the PES can be significantly enhanced by increasing the complexity of GP model kernels. All previous authors considered GP models of PES with simple, fixed kernels. The accuracy of such models was increased by increasing the number of potential energy points. However, the numerical difficulty of training a GP model scales with the number of potential energy points $n$ as ${\cal O}(n^3)$ and the evaluation time of a GP model, once trained, scales as ${\cal O}(n)$. It is therefore necessary to develop approaches that produce more accurate GP models with smaller $n$. 
The present work demonstrates an interpolation approach that can be used to enhance the accuracy of the PES model with a fixed number of energy points by building composite kernels with increased complexity. In particular, we propose an algorithm to enhance the accuracy of the PES through optimization in the space of kernels by varying random distributions of {\it ab initio} points with fixed $n$. Our final result illustrates that GP models of PES for polyatomic systems can be constructed to be accurate both within and {\it outside} the range of the given potential energy points.  

%Such models can be used to explore the global landscape of complex PES given a small number of randomly selected electronic structure calculations and obtain accurate global PES with random distributions of potential energy points. 

This work builds on the demonstration that the predictive power of GP models can be enhanced by increasing the kernel complexity using the Bayesian information criterion (BIC) as a metric \cite{extrapolation-1,extrapolation-2,extrapolation-3}. The kernel selection algorithm based on the maximization of the BIC was used in Ref. \cite{extrapolation-3} to extrapolate quantum properties of lattice spin systems across quantum phase transition lines. Here, we apply the same algorithm to construct GP models of globally accurate PES and extend it to illustrate that further improvement of accuracy can be achieved by optimization in the space of kernels by varying the training data distributions. 
%We first consider the interpolation problem, as in previous work \cite{gp-1,gp-2,gp-3,jie-jpb,gp-for-PES-2,gp-for-PES-3,gp-for-PES-4,gp-for-PES-5,gp-for-PES-6,gp-for-PES-7,gp-for-PES-8,gp-for-PES-9,gp-for-PES-10}, and show that the kernel selection algorithm advocated in Refs. \cite{extrapolation-1,extrapolation-2,extrapolation-3} can be used to increase dramatically the accuracy of the PES fits.  We then show that the algorithm of Refs. \cite{extrapolation-1,extrapolation-2} can also be used to obtain the physical fit of the PES outside the range of the given potential energy points. 
We illustrate the approach by constructing the six-dimensional PES for H$_3$O$^+$ computed in Ref. \cite{h3o+}. 
For the interpolation problem, we show that this algorithm produces a global six-dimensional PES for H$_3$O$^+$ with the  root mean square error (RMSE) $65.8$ cm$^{-1}$, using as input 500 randomly selected potential energy points in the energy range between zero and $21,000$ cm$^{-1}$.  To illustrate extrapolation, we produce the PES at high energies based on the energy points at low energies. We show that one can obtain an accurate global fit of the PES extending to $21,000$ cm$^{-1}$ based on 1500 potential energy points at energies below $10,000$ cm$^{-1}$.

\section{Composite kernels for GP models of PES}

The approach adopted here was described in Refs. \cite{extrapolation-1,extrapolation-2,extrapolation-3}. In brief, a Gaussian Process $y(\bm x)$ can be considered as a limit of 
a Bayesian neural network with an infinite number of hidden nodes and Gaussian priors for the NN parameters \cite{BML}. The inputs of the GP are the variables describing the internal coordinates of a polyatomic system, collectively denoted by the vector $\bm x = \left [ x_1, ..., x_N \right ]^\top$, where $N$ is the number of independent variables. The output  $y$ of the GP is the value of the potential energy.  At any value of $\bm x$, there is a normal distribution $P(y)$ of values $y$. 
When a GP is trained, the goal is to condition $P(y)$ in the entire configuration space by $n$ known values of the potential energy $\bm y = \left [y_1, ..., y_n \right ]^\top$ at $n$ points $\left [ \bm x_1, ..., \bm x_n \right ]^\top$ of the $N$-dimensional variable space. The mean of this conditional distribution at an arbitrary point $\bm x_\ast$ is given by \cite{BML,gp-book}
\begin{eqnarray}
\mu_\ast = \bm k_\ast^\top \bm K^{-1} \bm y,
\label{GP-mean}
\end{eqnarray}
where $\bm k_\ast$ is a vector with $n$ entries $k(\bm x_\ast, \bm x_i)$ and $\bm K$ is a square $n \times n$ matrix with entries $k(\bm x_i, \bm x_j)$. 
The quantities $k(\bm x_i, \bm x_j)$ are the kernels, which, given a particular choice of the GP prior \cite{BML,gp-book}, represent the covariance of the normal distributions of $y$ at $\bm x_i$ and at $\bm x_j$. Given the kernels, Eq. (\ref{GP-mean}) can be used to predict the value of the potential energy at the point $\bm x_\ast$. 
%Eq. (\ref{GP-mean}) produces a smooth and differentiable function of $\bm x_\ast$. 

To build a GP model, one assumes an analytical form for $k(\bm x, \bm x')$ with some unknown parameters. As covariances are expected to decrease with the distance between the points in the $N$-dimensional space of input variables, one typically assumes a kernel function that decays with $|\bm x - \bm x'|$. The choice of the kernel function is not unique. For example, an accurate model of a PES can be constructed with any of the following kernel functions: 
%\begin{eqnarray}
%k_{\rm C}({{\bm x}'}, {{\bm x}'}) = \ell
%\label{eqn:k_C}
%\end{eqnarray}
\begin{eqnarray}
k({\bm x}, {{\bm x}'})  =  {\bm x}^\top {{\bm x}'}
\label{eqn:k_LIN}
\end{eqnarray}
\begin{eqnarray}
%k_{RBF}({\bm x}, {{\bm x}'}) = \exp \left(-\frac{|{\bm x}- {{\bm x}'}|^2}{2\ell^2}\right)
k({\bm x}, {{\bm x}'}) = \exp \left(-\frac{1}{2}r^2({\bm x}, {{\bm x}'})\right)
\label{eqn:k_RBF}
\end{eqnarray}
\begin{eqnarray}
%k_{MAT}({\bm x}, {{\bm x}'})  = \left( 1 + \frac{\sqrt{5}|{\bm x}- {{\bm x}'}|^2}{\ell} +  \frac{5 |{\bm x}- {{\bm x}'}|^2}{3\ell^2}\right ) \exp\left ( -\frac{\sqrt{5}|{\bm x}- {{\bm x}'}|^2}{\ell}\right )
k({\bm x}, {{\bm x}'})  = \left( 1 + \sqrt{5}r({\bm x},{{\bm x}'}) +  \frac{5}{3}r^2({\bm x}, {{\bm x}'})\right )
\nonumber
\\
\times \exp\left ( -\sqrt{5}r({\bm x}, {{\bm x}'})\right )~~~~
\label{eqn:k_MAT}
\end{eqnarray}
\begin{eqnarray}
k({\bm x}, {{\bm x}'})  = \left ( 1 + \frac{|{\bm x}- {{\bm x}'}|^2}{2\alpha\ell^2} \right )^{-\alpha}
\label{eqn:k_RQ}
\end{eqnarray}
%\begin{eqnarray}
%k_{\rm PER}({\bm x}, {{\bm x}'})  =  \exp\left (- \frac{2\sin^2\left(\frac{\pi |{\bm x}- {{\bm x}'}|}{\alpha} \right)}{\ell^2}\right )
%\label{eqn:k_PER}
%\end{eqnarray}
where $r^2({\bm x}, {{\bm x}'}) = ({\bm x}- {{\bm x}'})^\top \times {\bm M} \times ({\bm x}-{{\bm x}'})$ and ${\bm M}$ is a diagonal matrix with $N$ parameters, one parameter for each dimension of ${\bm x}$. 
%The labels of the kernels denote `linear', `radial basis function', `Mat\'ern'  and `rational quadratic' kernels \cite{gp-book}. 
Once the kernel function is chosen, the parameters of the kernel function are found by maximizing the logarithm of the marginal likelihood \cite{BML,gp-book}
\begin{eqnarray}
\log {\cal L} = -\frac{1}{2}{\bm y}^\top  {\bm K} ^{-1}{\bm y} - \frac{1}{2}\log |\bm K | - \frac{n}{2} \log 2\pi. 
\label{log-likelihood-explicit}
\end{eqnarray} 
The marginal likelihood is generally used to quantify the quality of the model \cite{BML,gp-book}. However, in order to compare two GP models with kernel functions of different complexity, it is more suitable to use the Bayesian information criterion \cite{bic} defined as 
\begin{eqnarray}
{\rm BIC} = \log {\cal L} - \frac{1}{2} {\cal M} \log n,
\label{BIC}
\end{eqnarray} 
where $\cal M$ is the number of parameters in the kernel function. The second term in Eq. (\ref{BIC}) penalizes kernels with more parameters.

In all previous studies \cite{jie-jpb,gp-for-PES-2,gp-for-PES-3,gp-for-PES-4,gp-for-PES-5,gp-for-PES-6,gp-for-PES-7,gp-for-PES-8,gp-for-PES-9,gp-for-PES-10}, the GP models of PES were constructed with a fixed kernel function, such as one of the kernel functions (\ref{eqn:k_RBF}) - (\ref{eqn:k_RQ}). In the present work, we follow Refs. \cite{extrapolation-1,extrapolation-2} to build GP models of PES with composite kernels obtained by combining the functions (\ref{eqn:k_LIN}) - (\ref{eqn:k_RQ}). The algorithm to construct suitable kernels is schematically depicted in Figure 1. This kernel selection approach starts with four GP models trained with each of the kernel functions (\ref{eqn:k_LIN}) - (\ref{eqn:k_RQ}) separately. The BIC (\ref{BIC}) is computed for each of the models and the kernel function of the model with the largest BIC is selected as the preferred kernel $k_0$. The kernel $k_0$ is then combined with each of the four kernels (\ref{eqn:k_LIN}) - (\ref{eqn:k_RQ}) by forming linear combinations $c_0 k_0 + c_i k_i$ and products $c_0 k_0 \times k_i$, thus leading to eight new functions. Eight GP models are trained with each of these kernel functions by optimizing both the kernel parameters and the coefficients $c_i$. The kernel of the model with the largest BIC is selected as the new preferred kernel $k_0$ and the procedure is iterated. 

\begin{figure}[ht]
	\includegraphics[width=0.6\columnwidth]{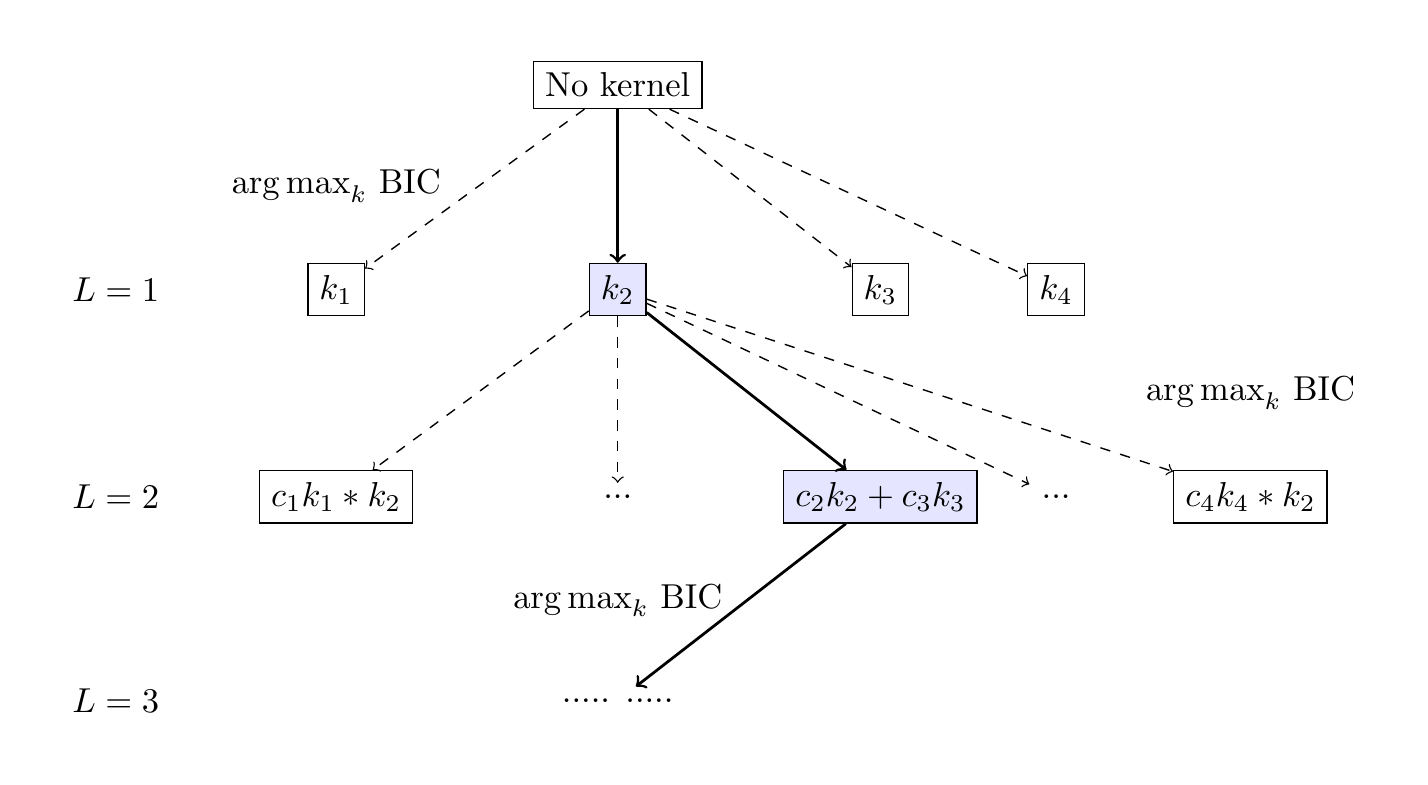}
	\caption{Schematic diagram of the composite kernel construction method. At each iteration,
	the kernel with the largest Bayesian information criterion (\ref{BIC}) is selected. {The labels in the boxes correspond to the kernel functions defined in (\ref{eqn:k_LIN})-(\ref{eqn:k_RQ})}.
	}
	\label{algorithm}
\end{figure}

Figure \ref{algorithm} depicts the kernel selection algorithm as a kernel search tree. We label the depth of the tree by $L$.
In the following sections, we will refer to the kernel functions in Eqs.  (\ref{eqn:k_LIN}) - (\ref{eqn:k_RQ}) as `simple' kernels, while the kernel functions at depth $L>1$ as composite kernels. 
As $L$ increases, the kernel functions become more complex and the kernel parameters become more difficult to optimize. It is therefore necessary to terminate the iterative kernel search algorithm at some value of $L$. The algorithm that selects the kernel with the largest BIC at each node of the tree is known as `greedy'. Note that the formulation of the kernel optimization as a search tree problem suggests other strategies, which may yield better kernels more efficiently. We leave the search for such strategies to future work.

\begin{figure}[h!]
	\includegraphics[width=0.6\columnwidth]{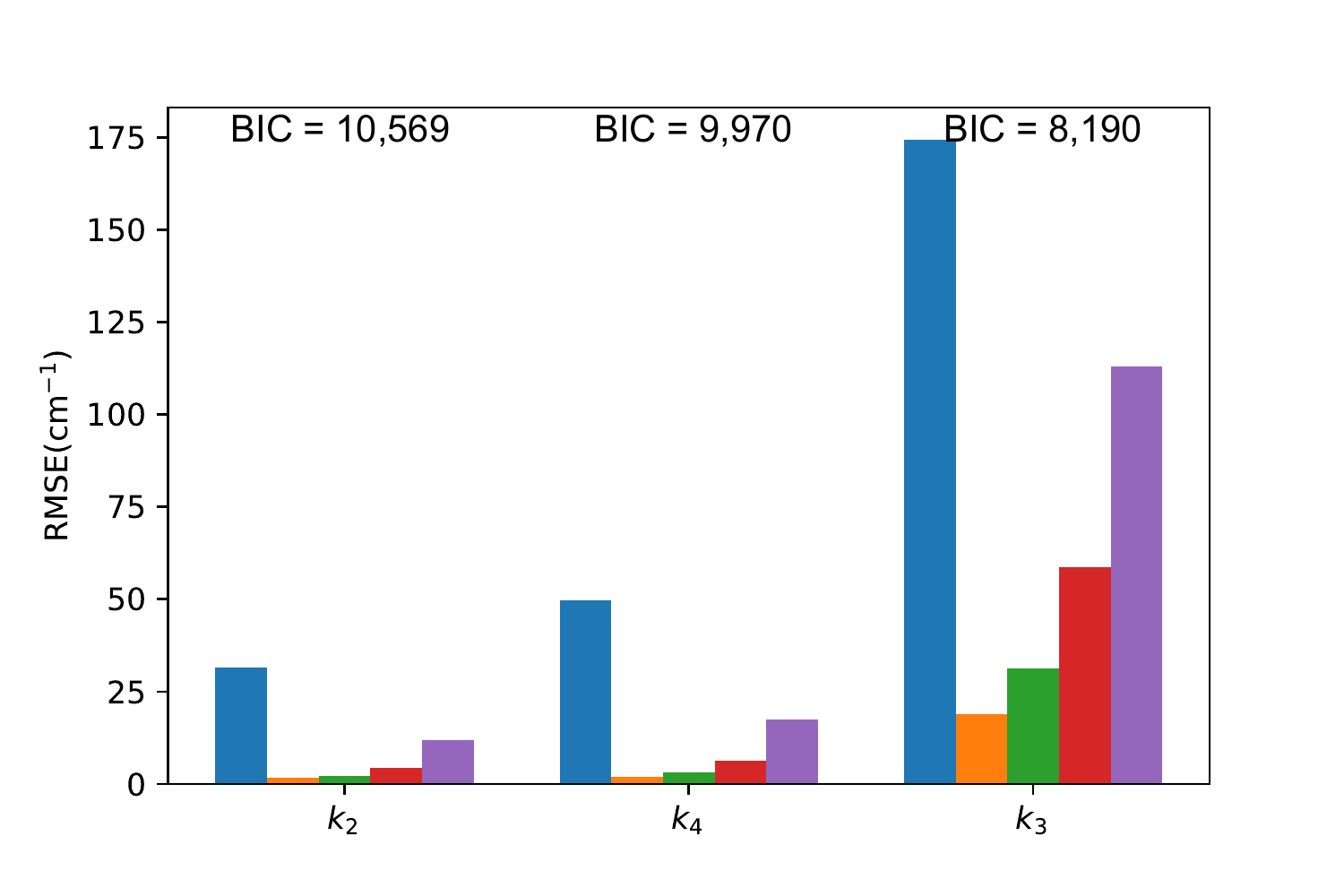} 
	\caption{RMSE for PES interpolation models with different kernels at kernel complexity level $L=1$. These interpolation models are trained by 1500 points selected from different energy intervals:  $[0, 5000]$ cm$^{-1}$ (orange),  $[0, 7000]$  cm$^{-1}$ (green),  $[0, 10000]$ cm$^{-1}$ (red), $[0, 15000]$ cm$^{-1}$ (purple) and $[0, 21000]$ cm$^{-1}$ (blue) .  The RMSEs are calculated using the remaining energy points in the corresponding energy interval in the set from Ref. \cite{gp-for-PES-10}.   The BIC listed is calculated for the models covering the energy range from 0 to 21000 cm$^{-1}$. The BIC for the other models follows the same trend. 
 }
	\label{interpolation-bic}
\end{figure}

\section{Interpolation of PES with composite kernels}

In previous work \cite{gp-1,gp-2,gp-3,jie-jpb,gp-for-PES-2,gp-for-PES-3,gp-for-PES-4,gp-for-PES-5,gp-for-PES-6,gp-for-PES-7,gp-for-PES-8,gp-for-PES-9,gp-for-PES-10}, GP models of PES for polyatomic systems were constructed with one of the simple kernels  (\ref{eqn:k_LIN}) - (\ref{eqn:k_RQ}). The focus has been on improving the accuracy of GP models by increasing the number of potential energy points in the training set $\bm y$ and selecting an optimal distribution of these points in the configuration space. 
In this section we show, that the accuracy of the GP interpolation models of PES can be significantly enhanced by increasing the complexity of kernel functions as described in the previous section. 

As an illustrative example, we consider the six-dimensional PES for H$_3$O$^+$. The potential energy for this system was computed using a combination of two high-level {\it ab initio} methods
(79 213 points with CCSD(T)-F12/aVQZ and 2284 points with MRCI/aVTZ) in Ref. \cite{h3o+}. Ref. \cite{gp-for-PES-10} presents a series of GP models obtained by interpolation of different numbers (from $n=500$ to $n=10,000$) of the {\it ab initio} points from Ref. \cite{h3o+} in the energy range below $60$ kcal/mol ($\sim 21,000$ cm$^{-1}$). To improve the accuracy of the PES fits, the inputs to the GP model were expressed as the Morse variables. It was found that the Morse variables produce significantly smaller RMSEs \cite{gp-for-PES-10}. The GP models were constructed with the radial basis function kernel (\ref{eqn:k_RBF}), a popular choice for GP regression. The RMSE of the GP model trained by 500 {\it ab initio} points was found to be 238.74 cm$^{-1}$. This error reduces to 125.76 cm$^{-1}$ for 1000 training points. 
We use these independent results as a reference for the present results. 

In order to illustrate the power of the BIC as a PES model selection criterion, we begin by training GP interpolation models with $1500$ energy points in five energy intervals ($[0,5000]$, $[0,7000]$, $[0,10000]$, $[0,15000]$, and $[0,21000]$ cm$^{-1}$) using different simple kernels given by Eqs. (\ref{eqn:k_LIN}) - (\ref{eqn:k_RQ}). The training points are selected randomly from the same set of the {\it ab initio} data as used in Ref. \cite{gp-for-PES-10}.
Figure \ref{interpolation-bic} shows the RMSE of the resulting models computed with all {\it ab initio} points, except the ones used for training, in the same energy interval as covered by the training distribution. Specifically, we use 4741 points in the energy interval $[0,5000]$ cm$^{-1}$, 8738 points  $\in [0,7000]$ cm$^{-1}$, 14872 points $\in [0,10000]$ cm$^{-1}$, 24174 points $ \in [0,15000]$ cm$^{-1}$ and 31124 points $ \in [0,21000]$ cm$^{-1}$ to compute these RMSEs. 
These results correspond to the output of the first layer ($L=1$) in the kernel search tree shown in Figure \ref{algorithm}.  As illustrated by Figure \ref{interpolation-bic}, the accuracy of the PES models is dramatically enhanced for the kernel with the largest BIC. This suggests that the kernel selection for GP models, even with simple kernels, can be guided by the value of the BIC. This is a valuable result because it suggests that the kernel selection for GP models of PES can be automated, even when composite kernels cannot be constructed due to numerical limitations.

\begin{figure}[h!]
	\includegraphics[width=0.6\columnwidth]{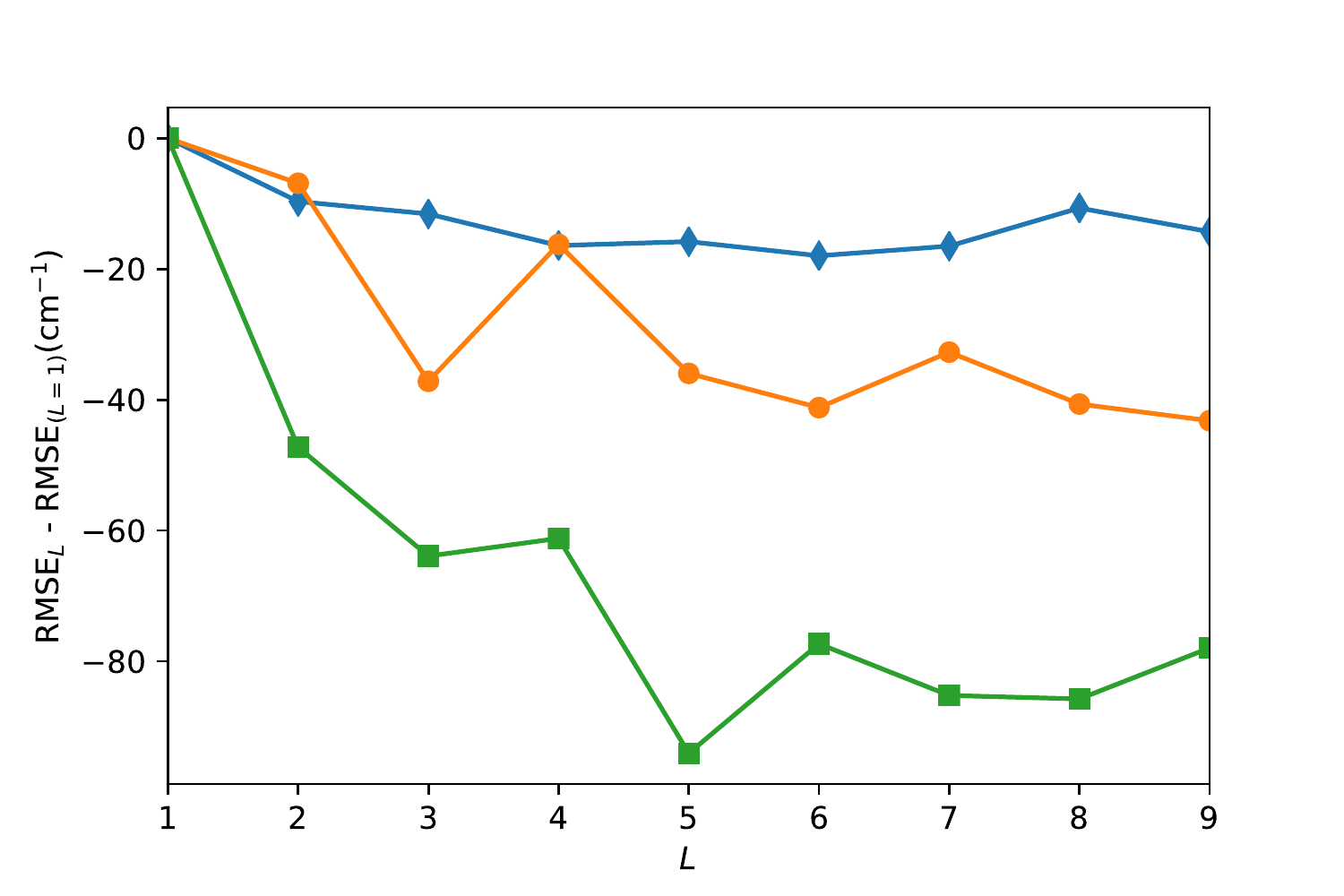} \\
	\includegraphics[width=0.6\columnwidth]{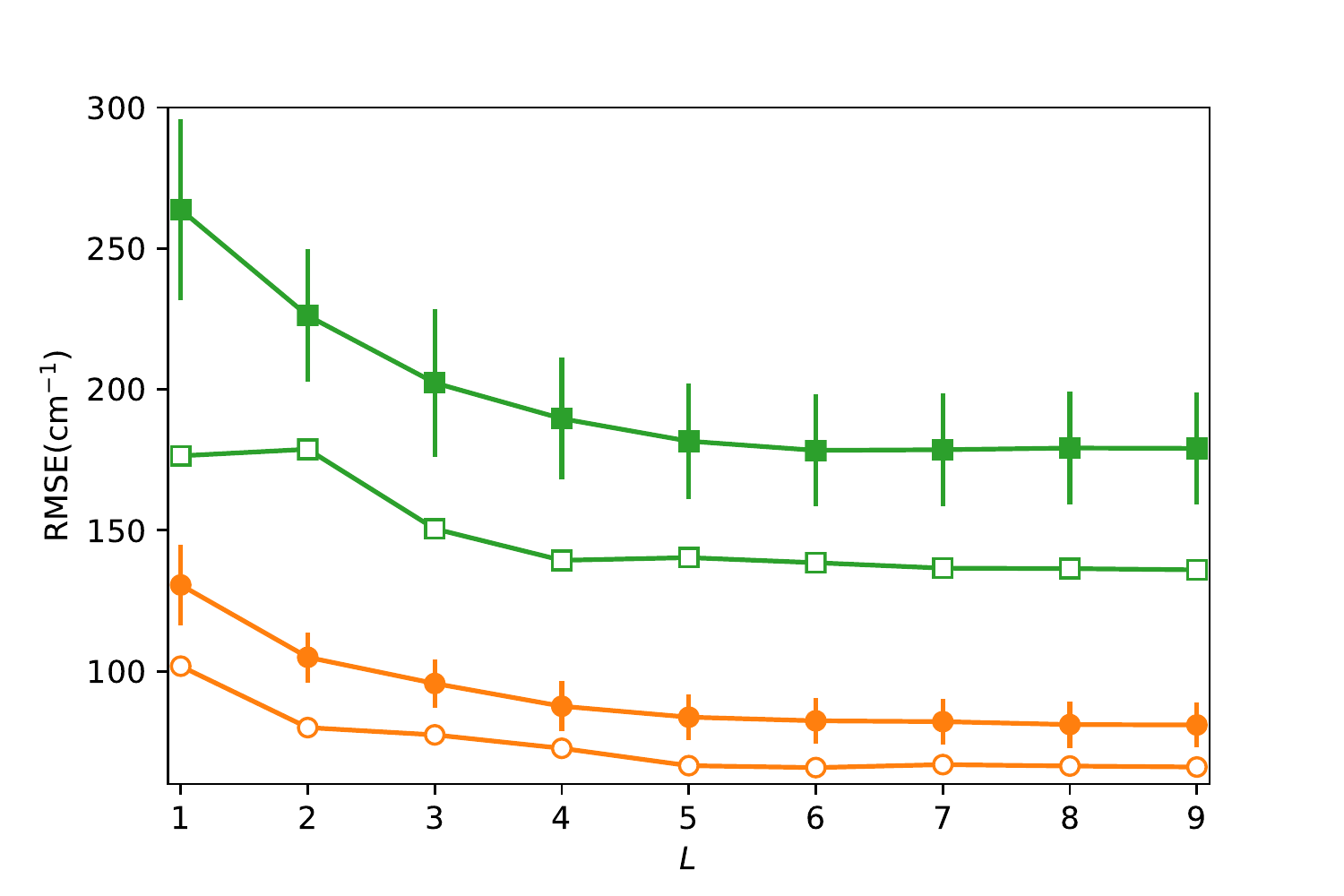}
	\caption{Upper panel: The dependence of the RMSE for the GP interpolation models of global six-dimensional PES trained by a single, randomly selected distribution of $n=300$ energy points (squares), $n=500$ energy points (circles) and $n=1,000$ energy points (diamons) on the layer $L$ of the kernel complexity illustrated in Figure 1.
The PES spans the energy interval $[0, 21000]$ cm$^{-1}$. 	
	The RMSE values for the models at the $L=1$ level are 260.89 cm$^{-1}$ (300 points), 116.26  cm$^{-1}$ (500 points), 45.91 cm$^{-1}$ (1,000 points).
Lower panel: Same as in the upper panel, but trained by 100 different distributions of energy points with fixed $n$: squares  -- $n=300$; circles -- $n=500$. The vertical bars show the standard deviation of the RMSE for the model ensemble with 100 models, excluding outliers. The open symbols show the RMSE of the most accurate model in the entire ensemble for a given $L$.  
	}
	\label{interpolation}
\end{figure}

Figure \ref{interpolation} shows that further improvement of the PES model accuracy can be achieved by using kernels from deeper levels of the kernel selection tree, illustrating that composite kernels, if selected by the algorithm of Figure \ref{algorithm}, lead to more accurate models of PES. 
%Figure \ref{interpolation} shows the RMSE of the GP models with composite kernels obtained using the algorithm of Figure \ref{algorithm}.  
%in Ref. \cite{h3o+}, covering the energy range between zero and $XXX$ cm$^{-1}$. 
The RMSE in Figure \ref{interpolation}  is calculated using $31124$ {\it ab initio}  test points that are not used for training the models. The training points and the test points for the models in Figure \ref{interpolation} are sampled from the energy interval $[0,21000]$.
 The $x$-axis of Figure \ref{interpolation} depicts the depth of the layer of the binary tree in Figure \ref{algorithm} used for the construction of the composite kernel. The first point ($L=1$) represents the RMSE of the PES obtained with one simple kernel of Eqs.  (\ref{eqn:k_LIN}) - (\ref{eqn:k_RQ}) that yields the largest BIC. The upper panel shows the results for a given, fixed distribution of {\it ab initio} points for each value of $n$. The lower panel shows the results for ensembles of GP models corresponding to multiple, random distributions of $n$ training points. 
We use the interatomic distances as the variables of this PES, which should yield less accurate results than the Morse variables used in previous work \cite{gp-for-PES-10}. 
 This is done for two reasons: (1) simplicity; (2) in order to emphasize that kernel functions, if allowed to be flexible, can compensate for the loss of accuracy due to suboptimal sampling.

Figure \ref{interpolation}  illustrates that increasing the complexity of the kernel functions by the algorithm of Figure \ref{algorithm} enhances the accuracy of the resulting GP model of the PES to a great extent. For the fixed distributions (upper panel of Figure \ref{interpolation}), the RMSE decreases by a factor of 1.55 (300 points), 1.59 (500 points), and  1.64 (1000 points) as
as the composite kernel complexity changes from one to nine layers. 
The resulting RMSE for the PES model trained with 500 {\it ab initio} points is significantly smaller than the RMSE of the PES model obtained with 1000 training points in Ref. \cite{gp-for-PES-10} (cf., 73 vs 125.76 cm$^{-1}$), despite the smaller number of training points and less optimal coordinates.  Our result for the PES model trained with 1000 points yields the RMSE of 38 cm$^{-1}$ (vs 125.76 cm$^{-1}$ in the independent reference). As discussed in the next section (and illustrated in the lower panel of Figure \ref{interpolation}), this result can be further improved by varying the random distribution of training points. 

Since Eq. (\ref{log-likelihood-explicit}) involves the computation of the determinant of the $n \times n$ matrix $\bm K$, the time to train a GP model scales as ${\cal O}(n^3)$ and the time to evaluate Eq. (\ref{GP-mean}) scales as ${\cal O}(n)$. 
While this makes the application of GPs to problems requiring large $n$ prohibitively difficult, training a GP model with $n \leq 1000$ does not present any numerical difficulty. 
Ref. \cite{gp-for-PES-10}  estimates the time to train a GP model with $n = 500$ to be 49 seconds. This time increases to 48662 seconds for $n=10,000$. The model evaluation time is even more important for the application of GP models of PES in molecular dynamics because the dynamical calculations, whether classical or quantum, evaluate the potential energy at a large number of points in the configuration space. Ref. \cite{gp-for-PES-10}  estimates the time to evaluate a GP model with $n = 500$ to be 2.81 seconds. This time increases to 55.58 seconds for the GP model with $n = 10,000$. This underscores the importance of the results shown in Figure \ref{interpolation} for applications in molecular dynamics. 
Instead of improving the GP model by increasing $n$, 
the algorithm used here aims to increase the accuracy of the GP models by improving the predictive power of kernels by training multiple GP models with small $n$. 

% The number $n$ of training points required to build an accurate $N$-dimensional GP model increases with $N$ \cite{BML}. Therefore, the approach demonstrated here potentially paves the way for the application of GP regression for bigger molecular systems. 

\section{Composite kernel optimization by varying training data}

The algorithm used in the previous section begins with a random distribution of {\it ab initio} points (training points) and aims to find the optimal composite kernel leading to the largest value of the BIC. As illustrated, this algorithm yields accurate GP models of PES with a small number of training points $n$. As discussed above, when $n$ is small, the numerical difficulty of training a GP model is insignificant. This can be exploited to optimize composite kernels further by varying the distribution of training points with fixed $n$. 

The problem can be considered as a simultaneous optimization in the space of composite kernels and the space of distributions of training points. 
To illustrate that such an approach yields more accurate models, we begin by defining $100$ random distributions of {\it ab initio} points for $n = 300$ and $n=500$. Each of these distributions is then used as a training set to build GP models with composite kernels of different complexity levels. For each distribution, the final composite kernel selected by the BIC maximization algorithm is different. This creates a distribution of models, each with a different RMSE, for each value of $L$. The lower panel of Figure \ref{interpolation} shows the mean and the standard deviations of these distributions for each value of $L$. 
Figure \ref{interpolation} also shows the lowest value of the RMSE in the entire ensemble of models for each value of $L$. 

Figure \ref{interpolation} thus illustrates an algorithm that can be used to enhance the accuracy of GP models through optimization in the space of kernels illustrated in Figure \ref{algorithm} by varying the distributions of the training points. The simplest version of this algorithm involves training an ensemble of GP models with different distributions (of a small number of {\it ab initio} points) and simply selecting the kernel and the training distribution of the model with the largest value of the BIC. The comparison of the results in the two panels of Figure \ref{interpolation} shows that this algorithm yields significantly more accurate models of the PES (open symbols in the lower panel). The lower panel of Figure 3 also illustrates that this algorithm requires fewer layers of the kernel search tree to produce the most accurate GP models and that the lowest RMSE thus obtained is a slowly varying function of $L$ at $L>4$.  As kernel optimization requires more computational effort as $L$ increases, we recommend the algorithm illustrated in the lower panel of Figure \ref{interpolation}  as the preferred algorithm for constructing accurate models of PES.

\section{Extrapolation of PES}

\begin{figure}
        \includegraphics[width=0.45\textwidth]{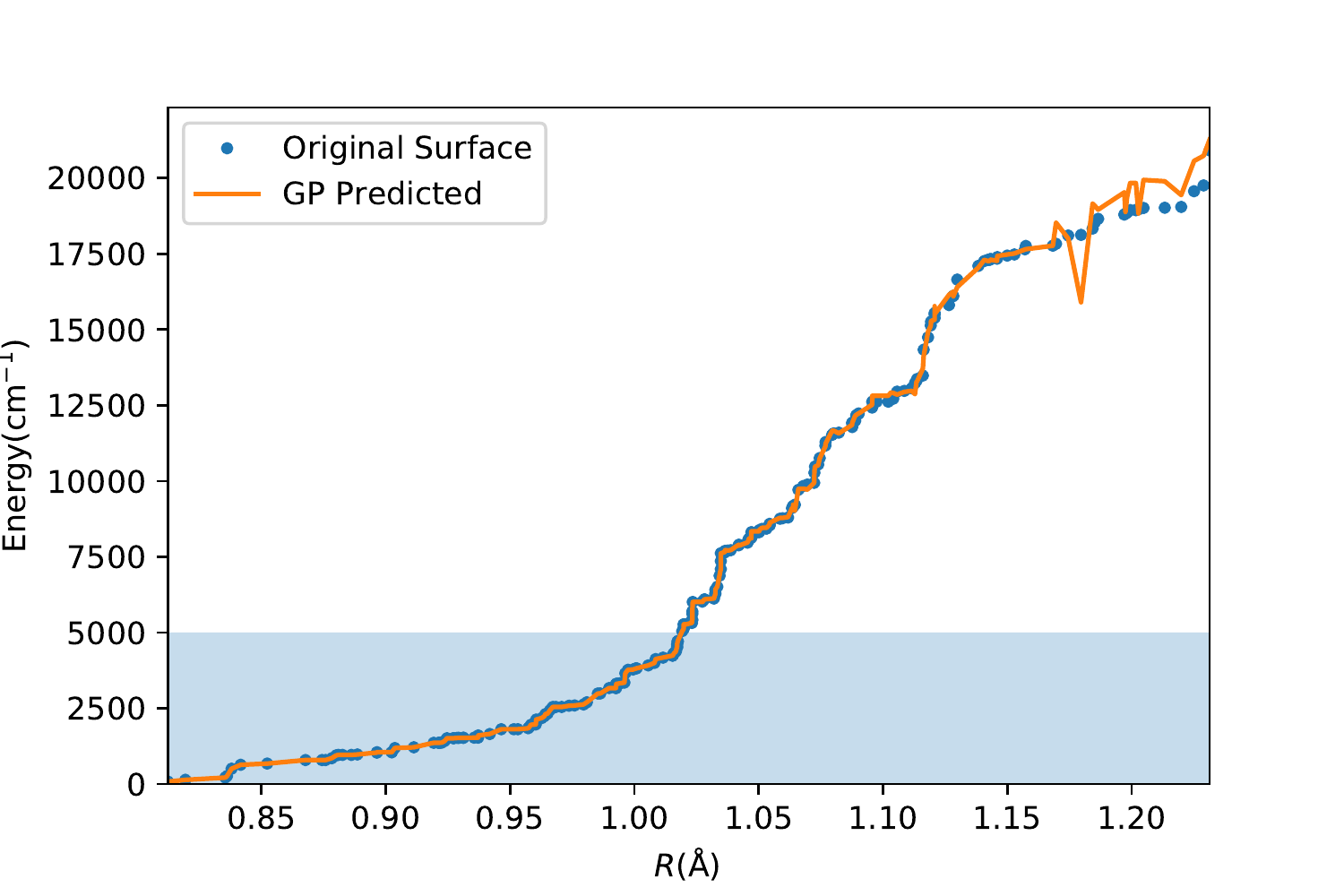}
        \includegraphics[width=0.45\textwidth]{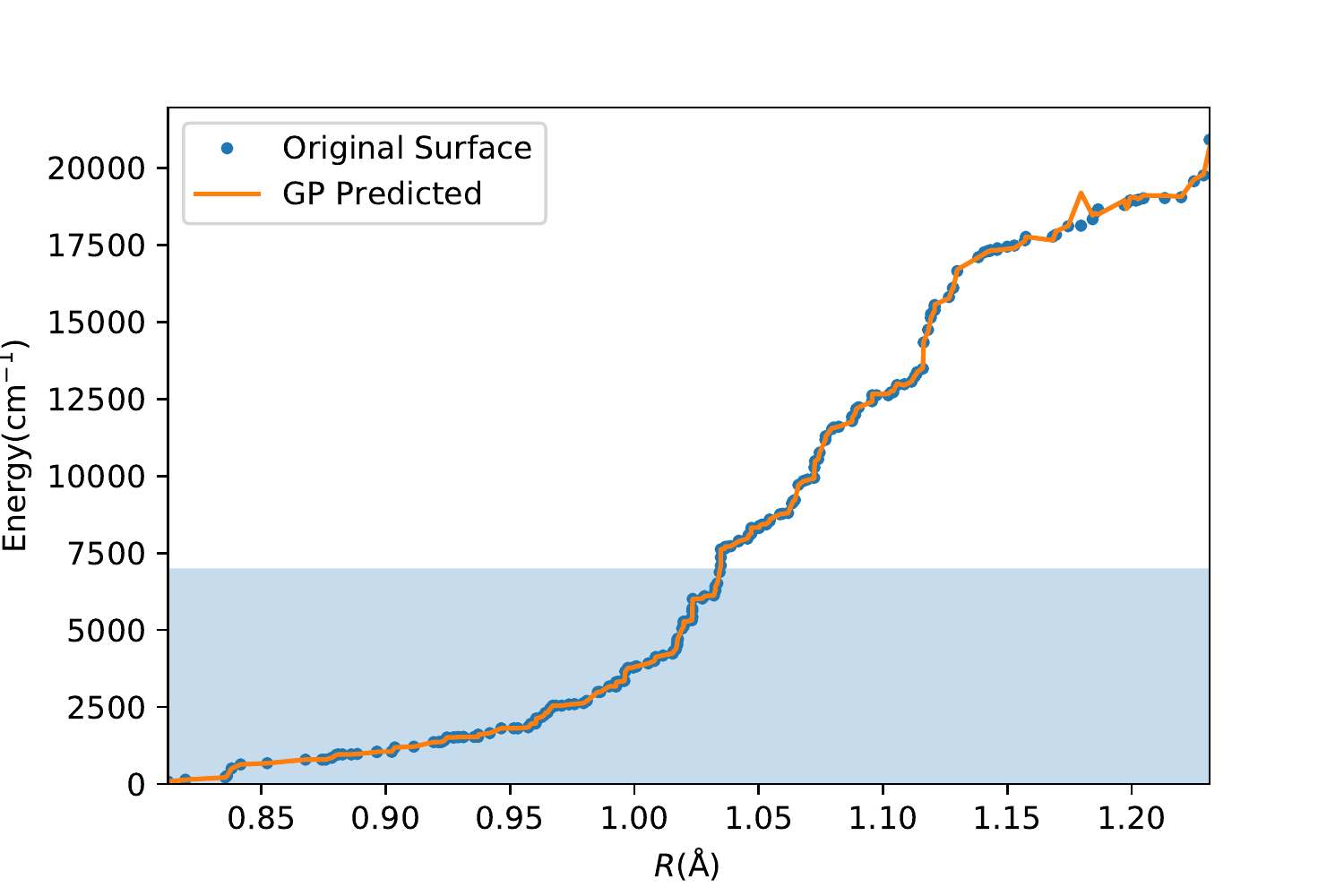}
        \includegraphics[width=0.45\textwidth]{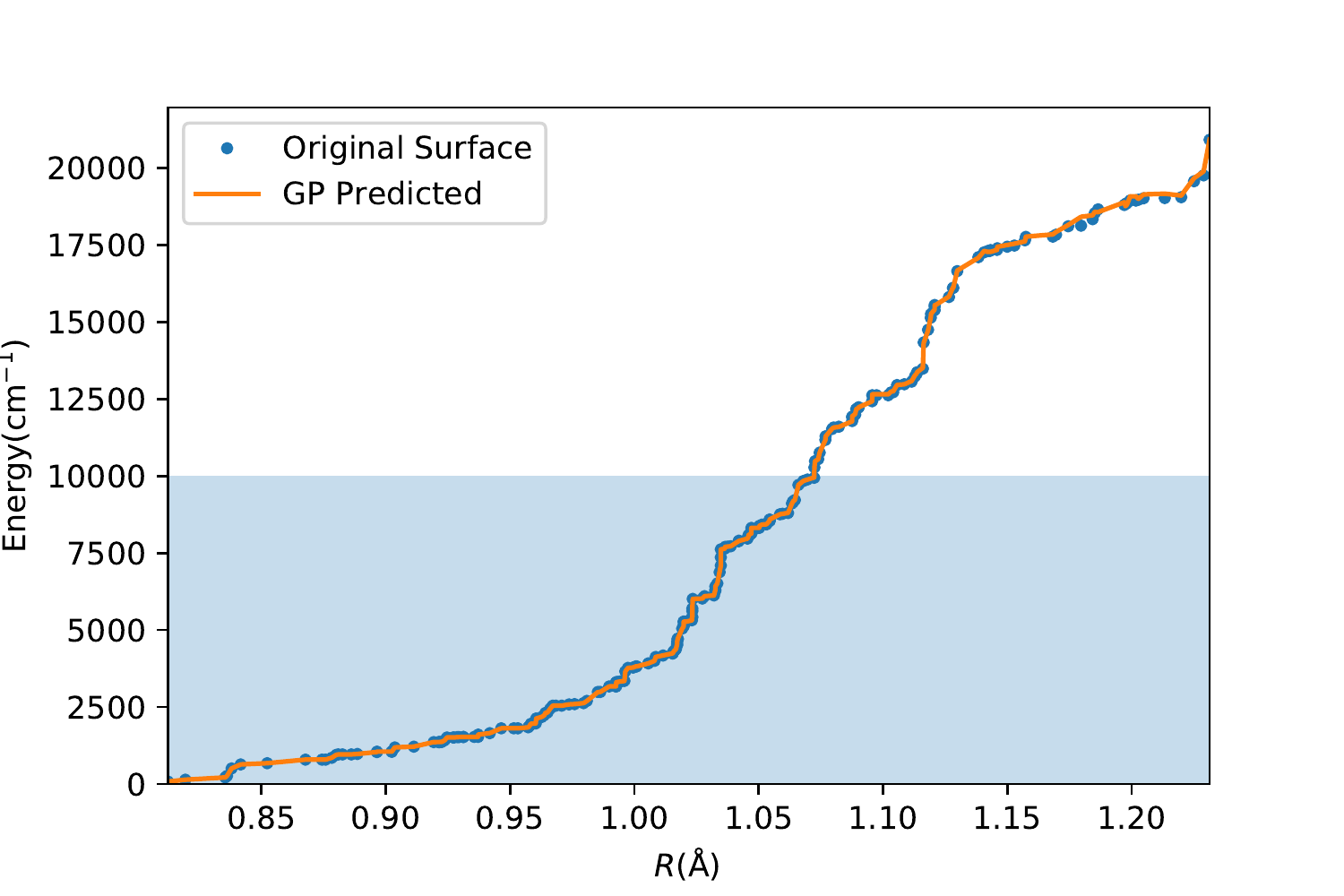}
        \includegraphics[width=0.45\textwidth]{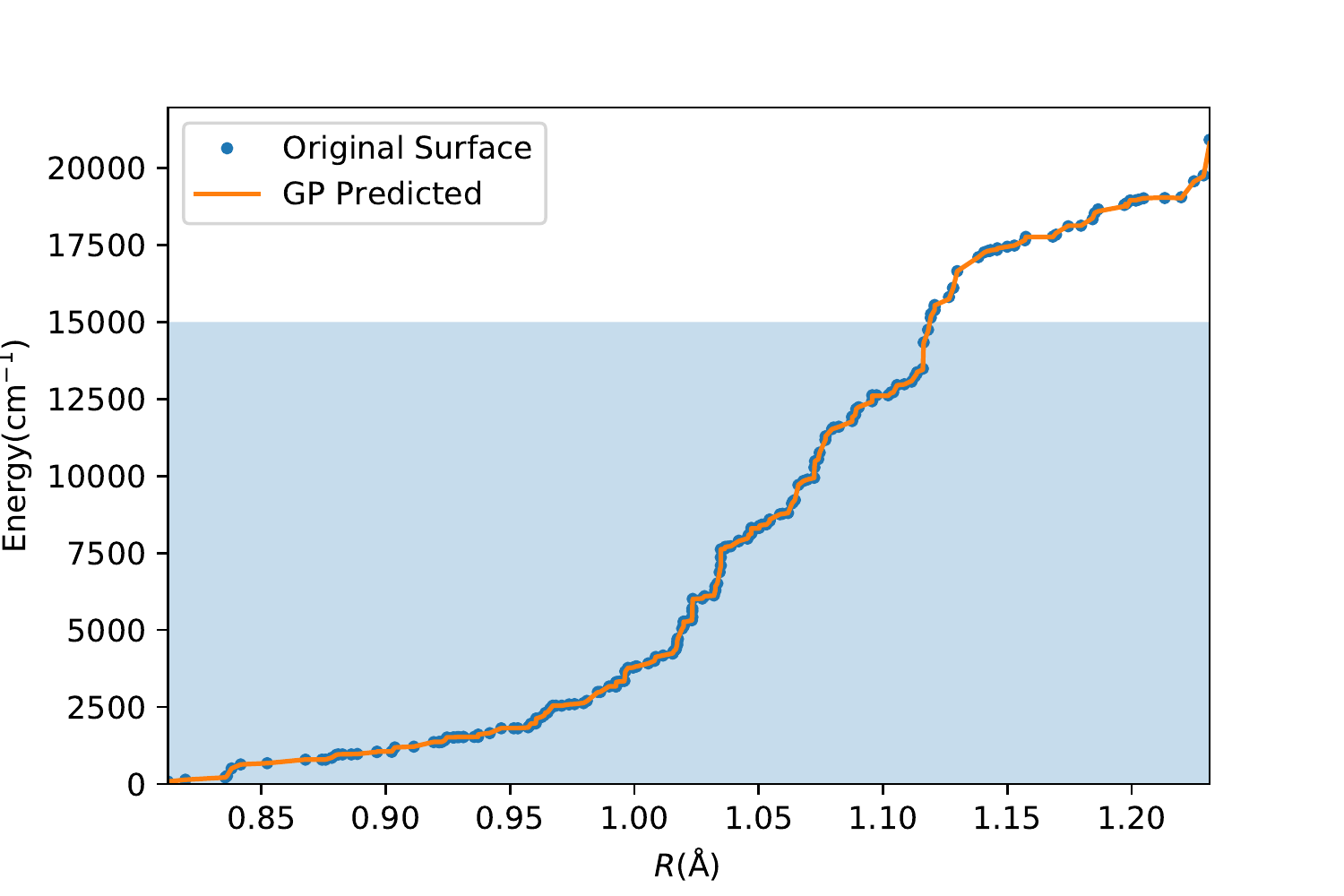}
    \caption{Comparison of the GP prediction (solid curves) with the original potential energy points (symbols) for OH$_3^+$. The GP models are trained by 1500 randomly selected {\it ab initio} points, strictly from the energy interval shown by the blue shaded region. To plot the energy, we separate the H$_2^+$ and OH fragments. The variable $R$ specifies the distance between the O atom and one of the H atoms in the H$_2^+$ fragment. At each value of $R$, we locate the energy point in the original set of {\it ab initio} points by varying the angles and/or the interatomic distances within the fragments. This energy point is compared with the GP prediction. 
}
\label{energy-extrapolation}
\end{figure}

\begin{figure}
        \includegraphics[width=0.45\textwidth]{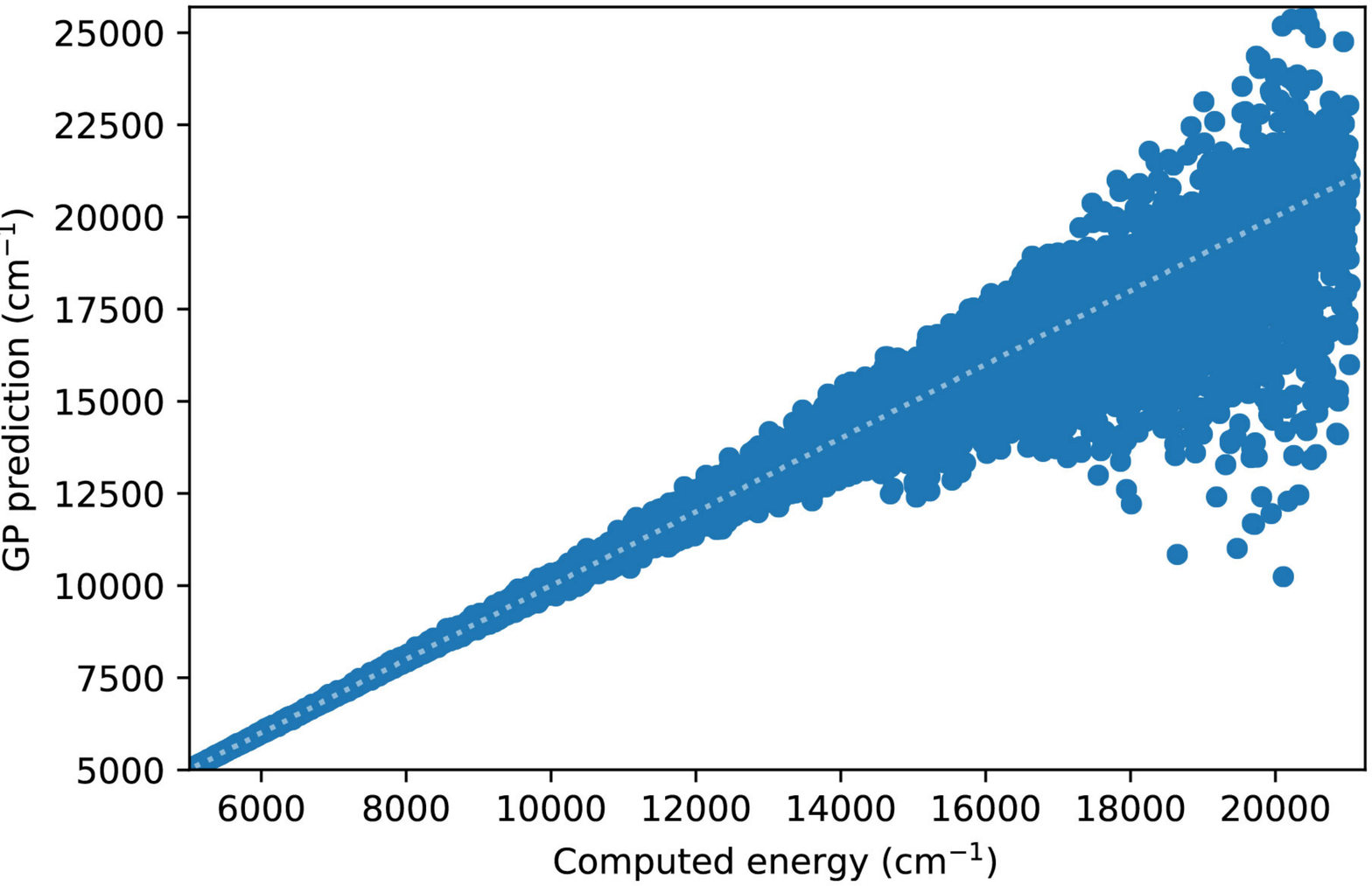}
        \includegraphics[width=0.45\textwidth]{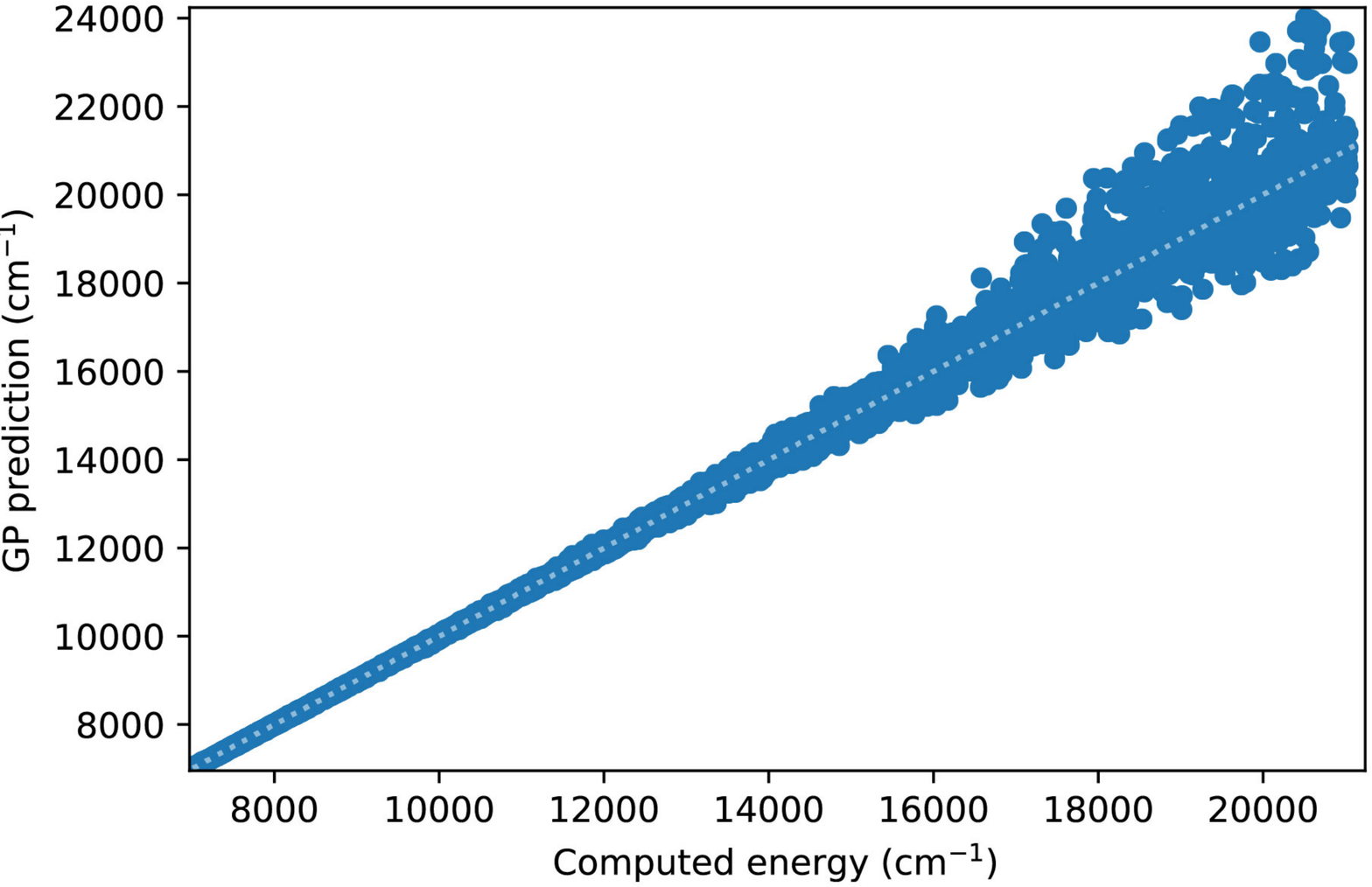}
        \includegraphics[width=0.45\textwidth]{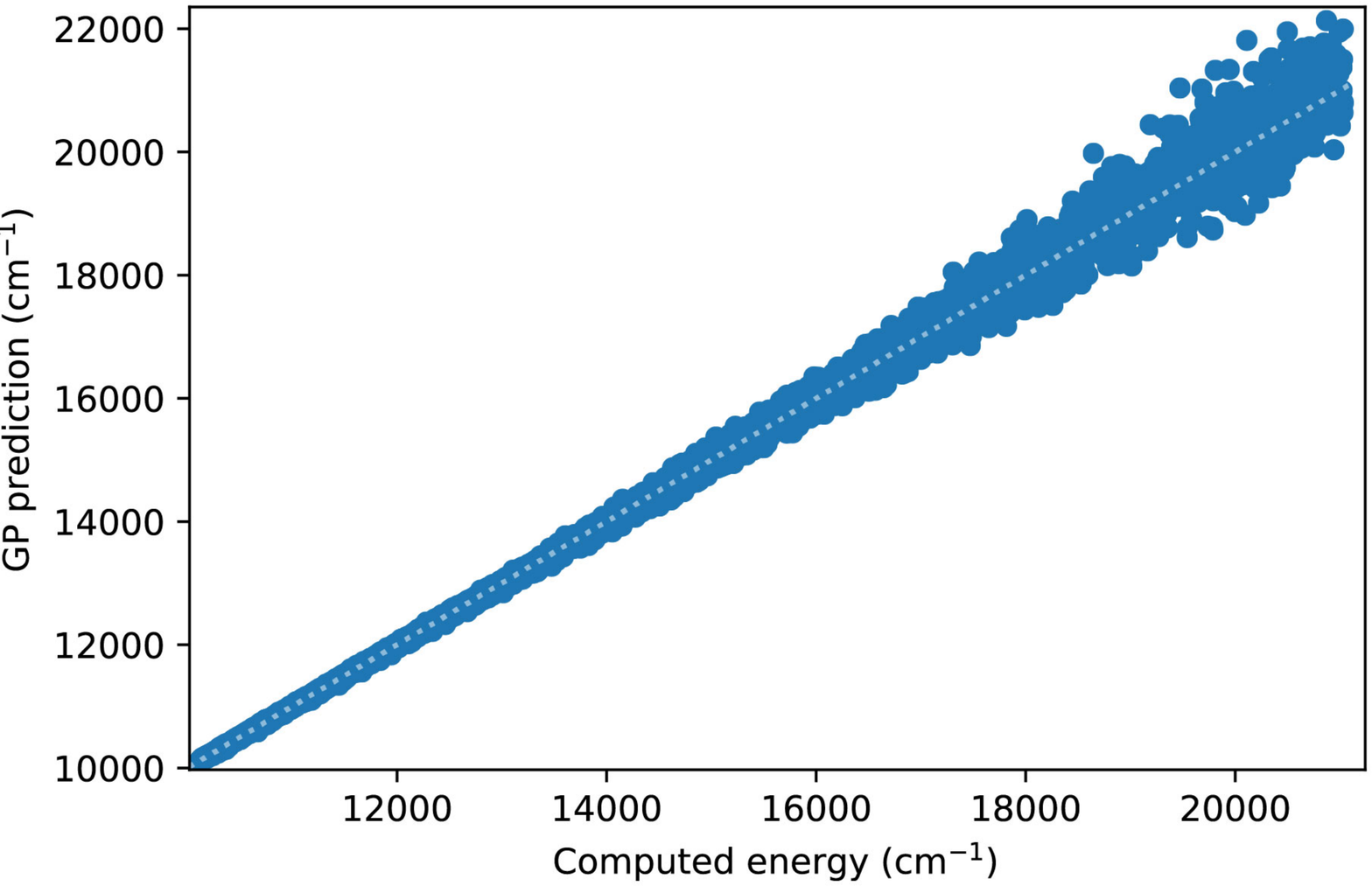}
        \includegraphics[width=0.45\textwidth]{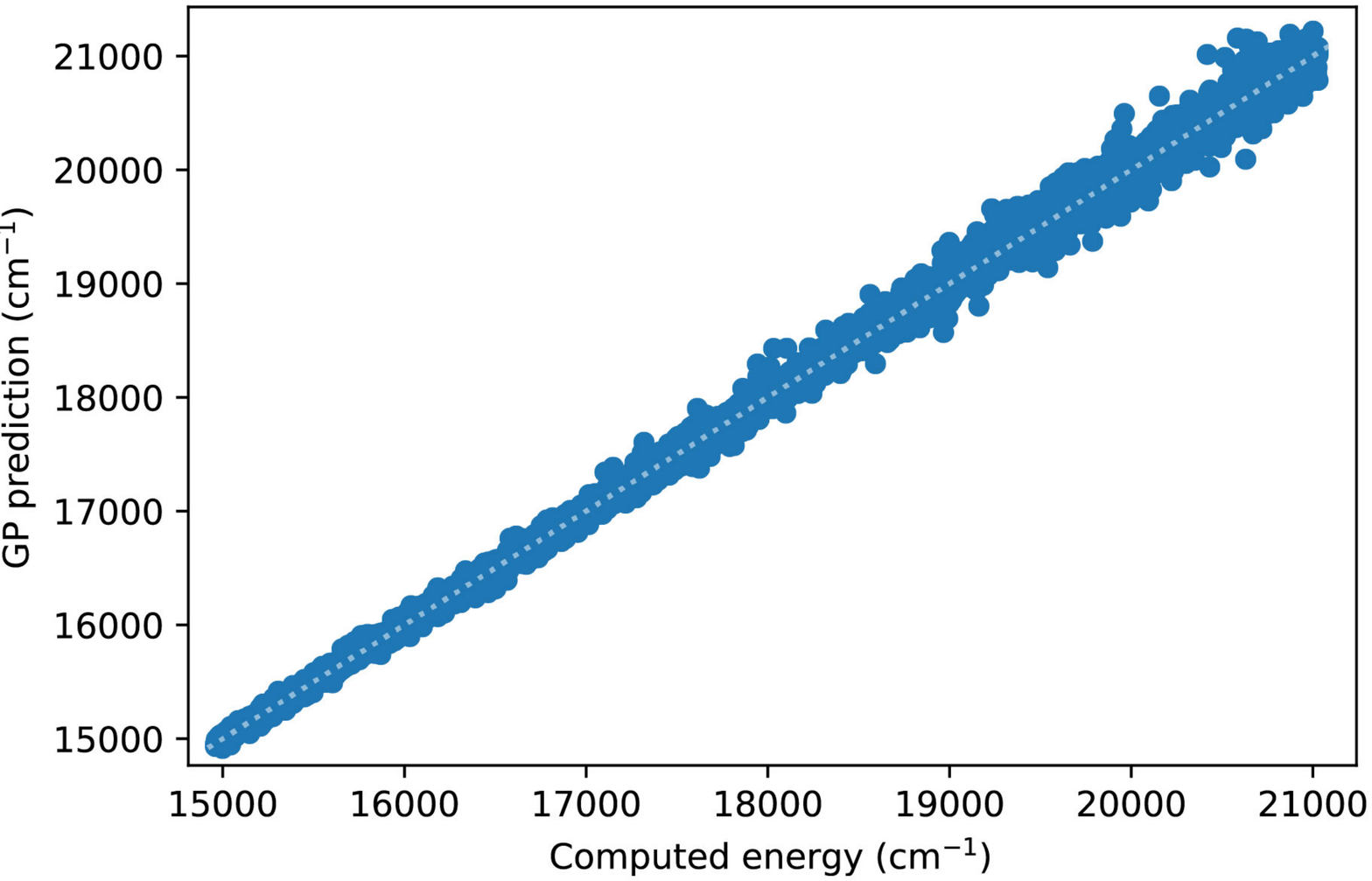}
    \caption{Comparison of the GP prediction with the original potential energy points for OH$_3^+$ at energies $E > E_{\rm threshold}$. The GP models are trained by random distributions of {\it ab initio} points at energies $E < E_{\rm threshold}$, with $E_{\rm threshold} = 5,000$ cm$^{-1}$ in the upper left panel, $E_{\rm threshold} = 7,000$ cm$^{-1}$ in the upper right panel, 
    $E_{\rm threshold} = 10,000$ cm$^{-1}$ in the lower left panel and $E_{\rm threshold} = 15,000$ cm$^{-1}$ in the lower right panel. The plots include all available {\it ab initio} points in the corresponding energy interval. 
 }
\label{scatter}
\end{figure}

To illustrate the feasibility of extrapolation by GP regression, we begin by constructing a series of GP models with $n=1500$ training points, selected at random but below a certain energy threshold $E_{\rm threshold}$. These models are then used to predict the value of the potential energy at energies $E > E_{\rm threshold}$.  The results are shown in Figure \ref{energy-extrapolation}. The GP models are obtained with kernels with the complexity level $L = 7$ of Figure \ref{algorithm}. For models with $n=1500$ training points, we stop the optimization of kernels at $L=7$, as further optimization becomes difficult due to the highly non-convex structure of the marginal likelihood.

\begin{figure}
        \includegraphics[width=0.75\textwidth]{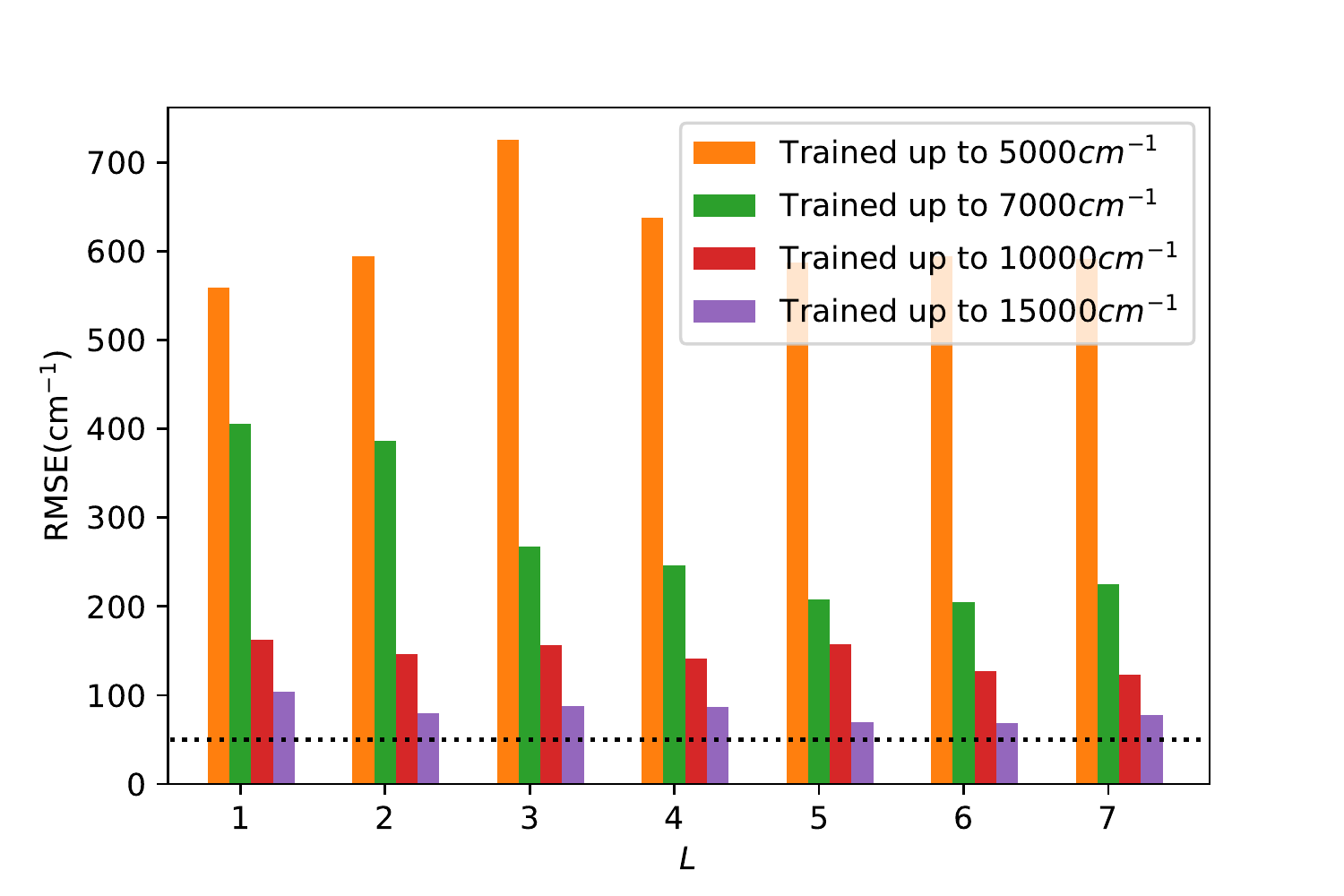}
     \caption{RMSE of the interpolation/extrapolation models trained by a fixed distribution of 1500 points selected from different energy ranges listed in the figure legend. The horizontal dotted line is at 50 cm$^{-1}$. The RMSEs are calculated using the energy points in the entire energy range between zero and $21,000$ cm$^{-1}$.}
\label{rmse-bar-plot}
\end{figure}

Figure \ref{energy-extrapolation} shows the energy of H$_3$O$^+$ at different separations between 
the H$_2^+$ and OH fragments. For each separation, we vary the relative angles of the fragments and their bondlengths to find an energy point in the set of {\it ab initio} points from Ref. \cite{h3o+}. The line in Figure \ref{energy-extrapolation}  is the prediction of the GP models, while the symbols show the {\it ab initio} points from Ref. \cite{h3o+}. The results of Figure \ref{energy-extrapolation} show that the GP models thus constructed remain accurate far beyond the energy  range of the training points. 

To illustrate quantitatively and non-ambiguously the accuracy of the models in the extrapolated energy range $E > E_{\rm threshold}$, we compare the GP prediction with the original {\it ab initio} point for each point in the entire set used in  Ref. \cite{gp-for-PES-10}. The results are shown in Figure \ref{scatter}. Note that the energy points in Figure  \ref{scatter} 
include all {\it ab initio} points from Ref. \cite{gp-for-PES-10} corresponding to all geometries of OH$_3^+$. Specifically, Figure  \ref{scatter}  includes 
26383 points above 5000 cm$^{-1}$; 22386 points above 7000 cm$^{-1}$;
16252 points above 10000 cm$^{-1}$; and 6950 points above 15000 cm$^{-1}$.
  These points thus sample the regions both within and outside the range of coordinates of the training points. 
%We emphasize that the plots in Figure  \ref{scatter}  include all $XXX$ {\it ab initio} points for comparison with our model predictions. 
The RMSE of the PES thus obtained are shown in Figure \ref{rmse-bar-plot}. As is clear from the numerical values of the RMSEs shown in Figure \ref{rmse-bar-plot}, most of the points in the scatter plots in Figure  \ref{scatter} are very close to the diagonal line, illustrating an agreement between the GP prediction and the original {\it ab initio} points. 

\begin{figure}
        \includegraphics[width=0.75\textwidth]{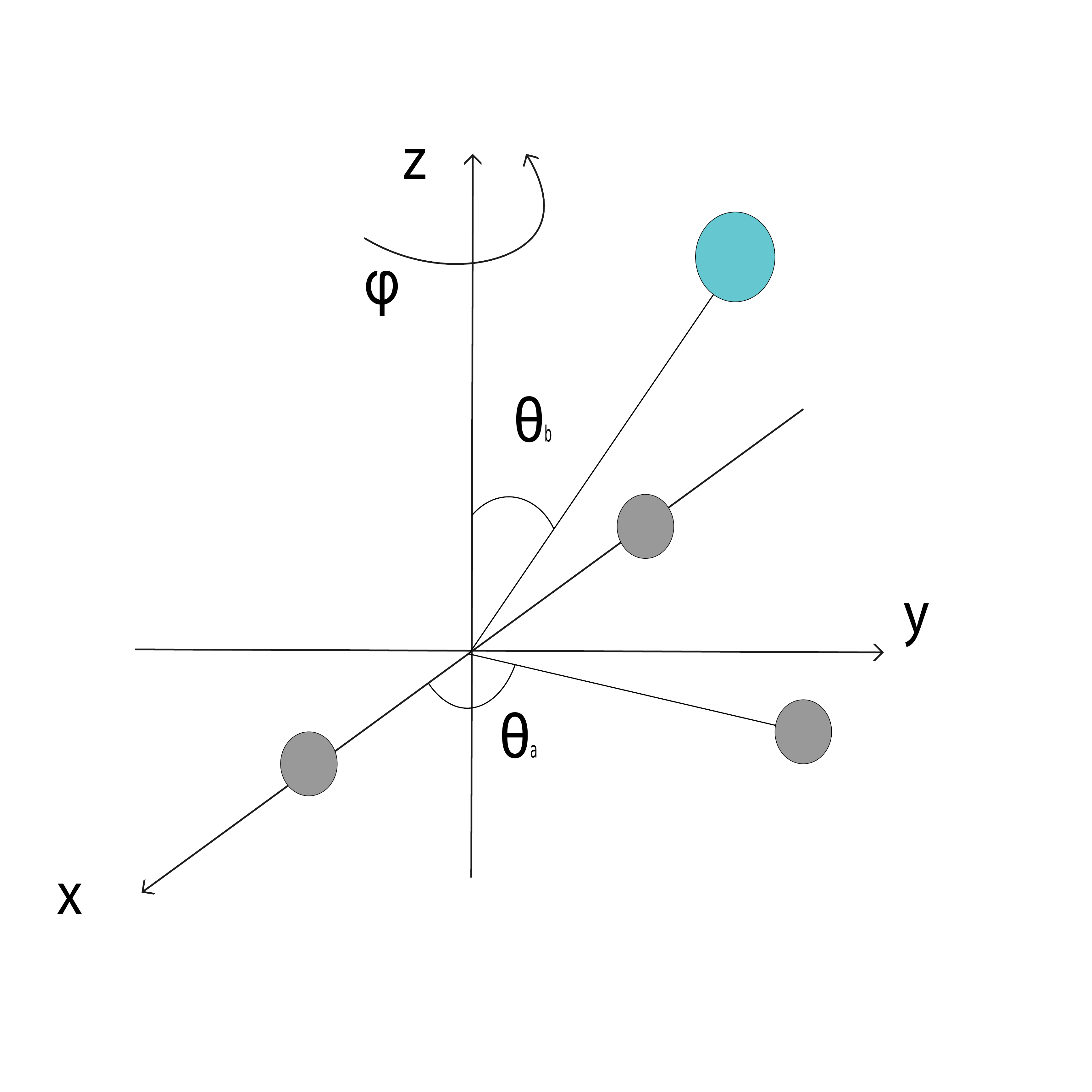}
     \caption{Schematic depiction of the variables used to describe the OH$_3^+$ complex in the present work. The smaller circles depict the H atoms, while the larger circle shows the position of the O atom.}
\label{geometry}
\end{figure}

We now illustrate the ability of GP models to extrapolate the PES as functions of the internal coordinates, to the part of the configuration space that is not sampled by {\it ab initio} data. 
To do so, we describe OH$_3^+$ by the internal coordinates shown in Figure \ref{geometry}. We fix the angles and the distance between the H atoms on the $x$ axes and plot the potential energy produced by the GP models as a function of the separation between H$^+$ and H$_2$O and the separation between O and H$_3^+$. The set of {\it ab initio} points does not describe this part of the configuration space. There is thus absolutely no information about this part of the configuration space in the training set of energy points. 

%The energy plotted in Figure \ref{pes-separation} is thus guaranteed to represent extrapolation. 

\begin{figure}
        \includegraphics[width=0.45\textwidth]{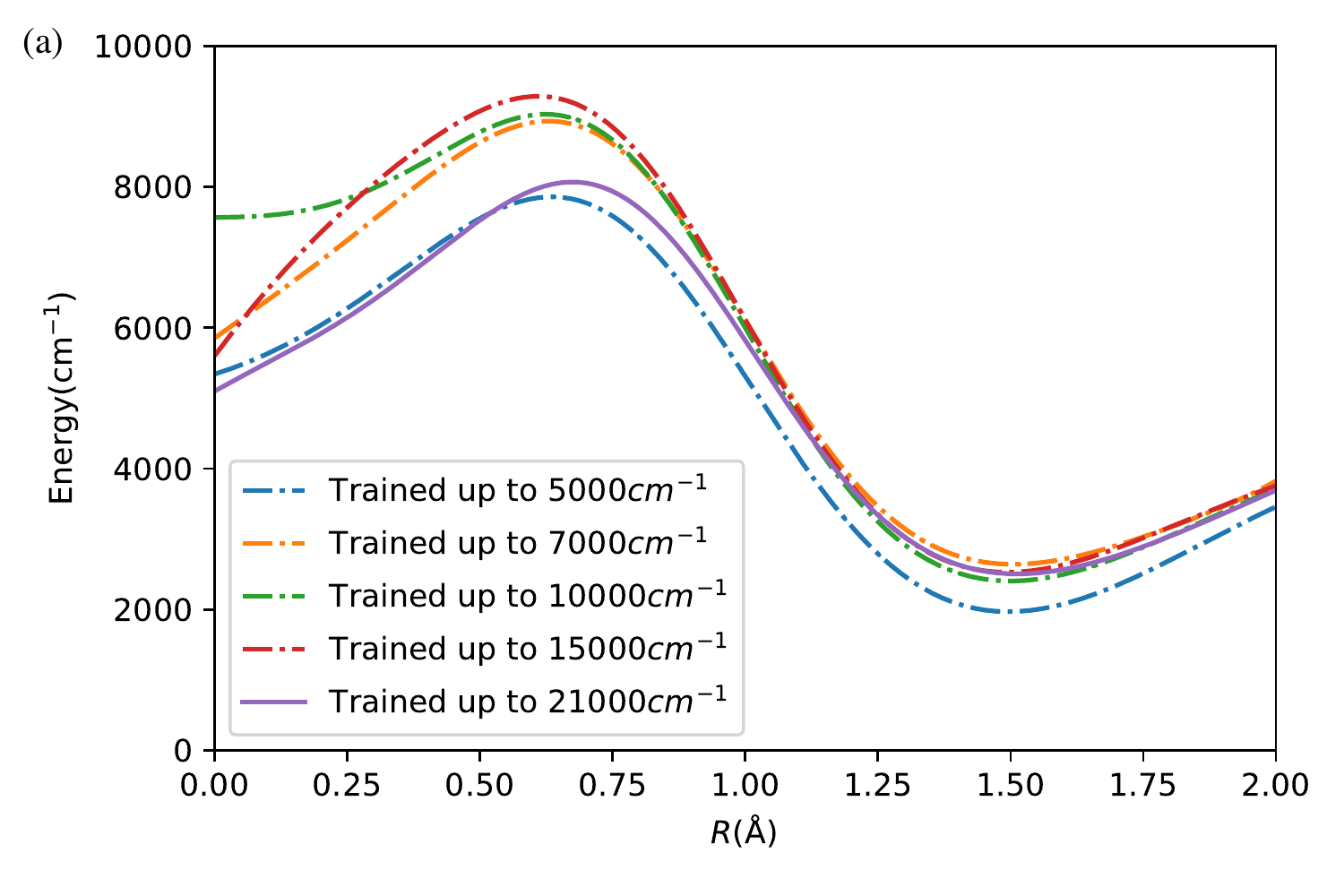}
        \includegraphics[width=0.45\textwidth]{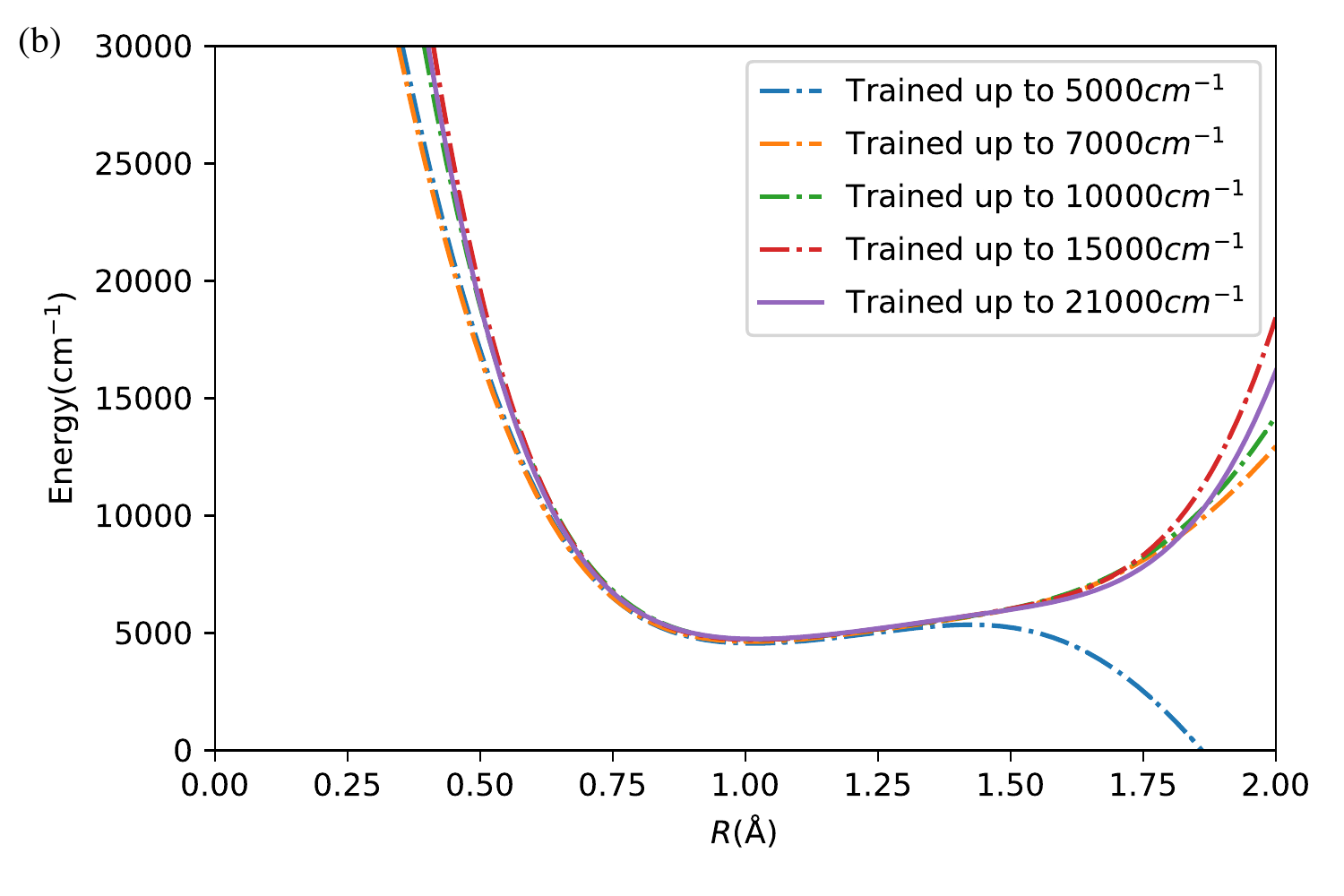}
        \includegraphics[width=0.45\textwidth]{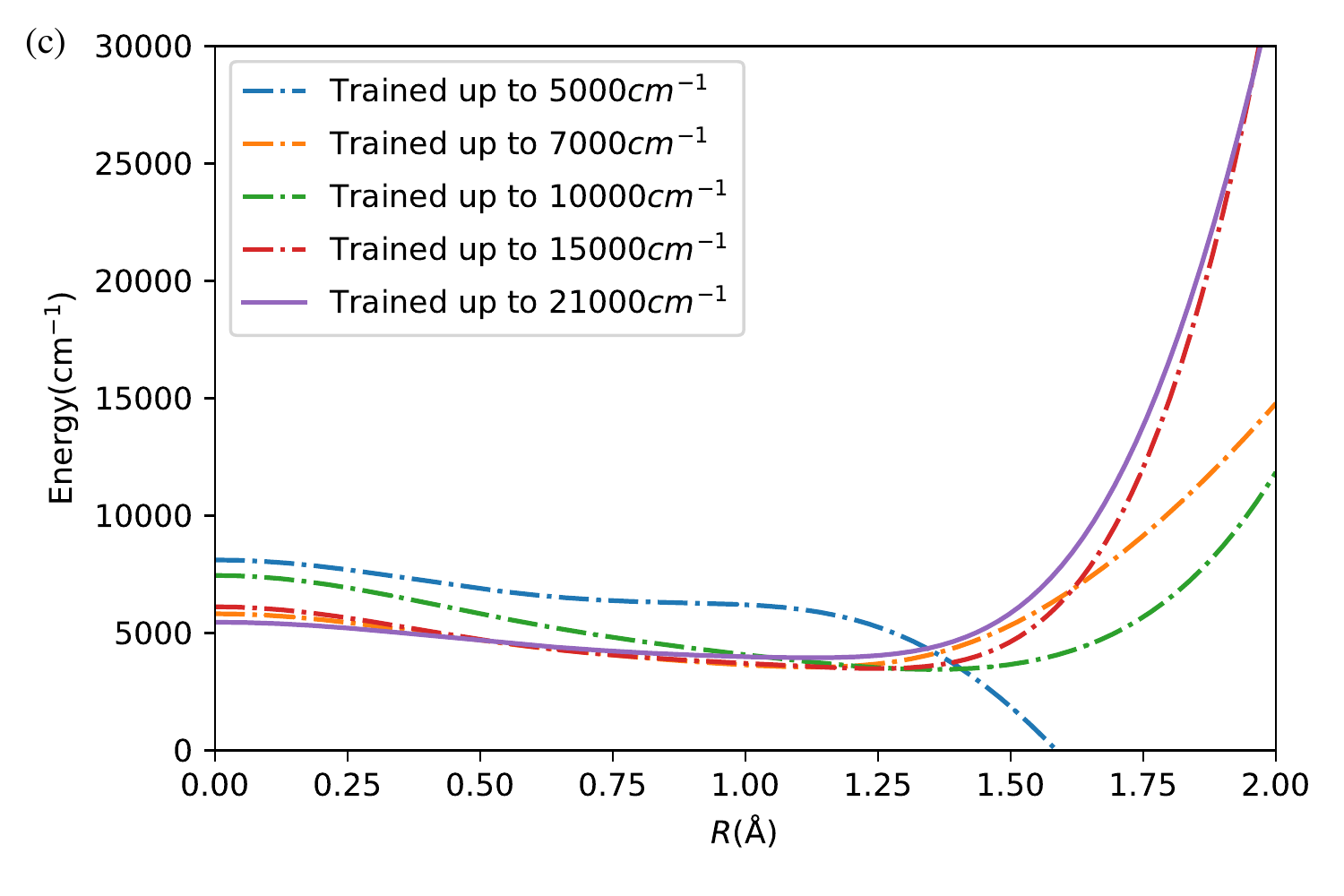}
        \includegraphics[width=0.45\textwidth]{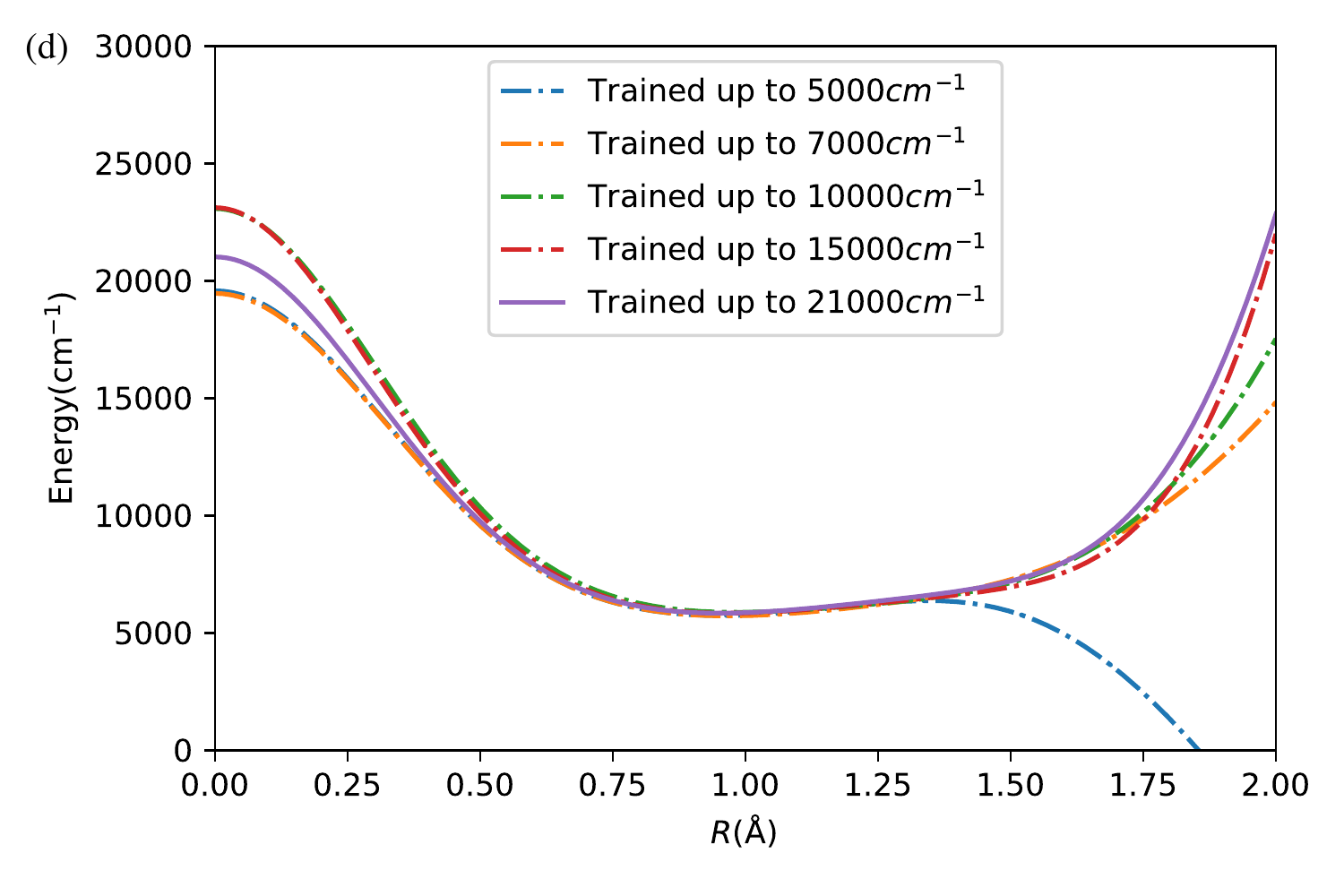}
        \includegraphics[width=0.45\textwidth]{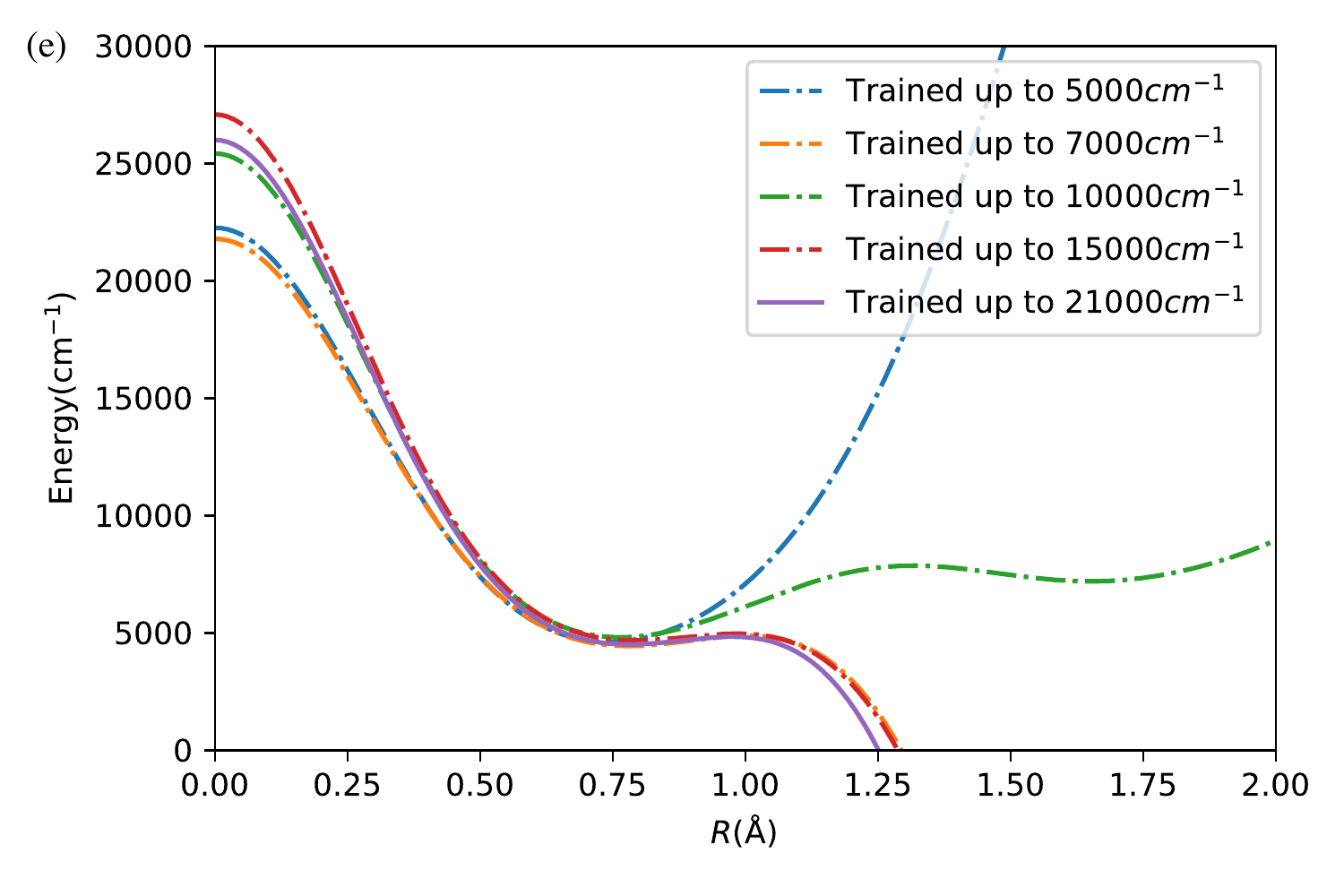}
        \includegraphics[width=0.45\textwidth]{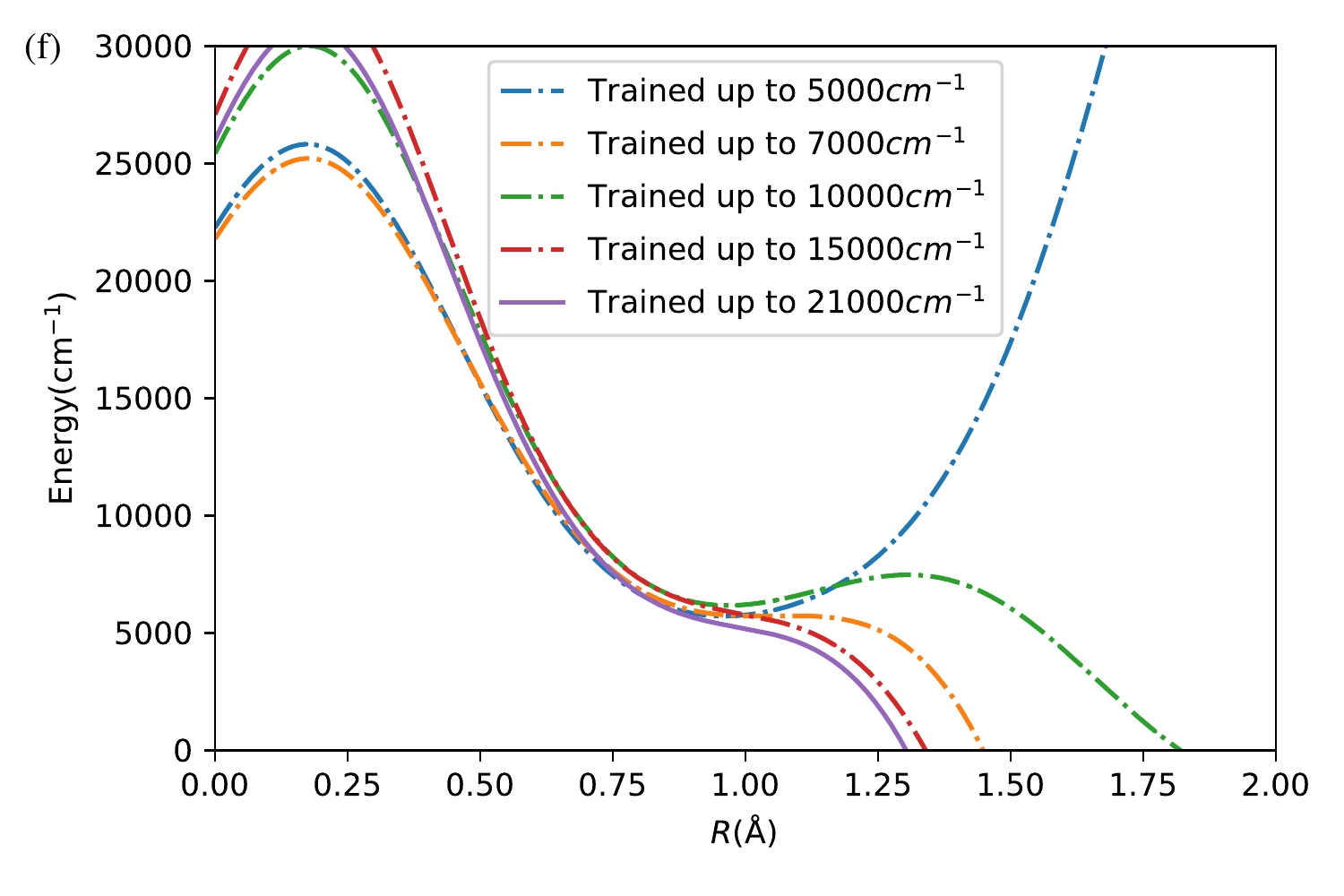}
    \caption{Six different cuts of the potential energy surface for $H_3O^+$ predicted by GP models. The two $H$ are fixed in x-axis with same distance between original. 
    The different panels correspond to the different angles  $(\theta_a,\theta_b,\phi)$ shown in Figure \ref{geometry}: panel (a)  -- {($\pi/2,\pi/3,\pi/3$)}; panel (b) --  {($\pi/2,0,0$)}; panel (c) -- 
 {($\pi/2,0,0$)}; panel (d) --  {($\pi/2,\pi/6,0$)}; panel (e) -- {($\pi/2,0,0$)}; panel (f) -- {($\pi/2,\pi/6,\pi/3$)}
  Panels (a)  through (d) show the energy as a function of the separation $R_b$ between H$^+$ and H$_2$O, while panels (e) and (f) show the energy as a function of the separation $R_c$ between  O and H$_3^+$.}
  \label{pes-separation}
\end{figure}

The different curves in Figure \ref{pes-separation}  correspond to predictions of five different GP models obtained, as before, by GP regression of the training points at energies below 5,000 cm$^{-1}$ (blue dot-dashed curve), 
7,000 cm$^{-1}$ (orange dot-dashed curve), 10,000 cm$^{-1}$ (green dot-dashed curve), 15,000 cm$^{-1}$ (red dot-dashed curve) and in the entire interval of energies (solid curve). 
While we have no {\it ab initio} data to validate the accuracy of the GP predictions in this part of the configuration space, we observe that five different GP models exhibit similar dependence on the atomic coordinates, except in a few cases,  where the models trained by low-energy points show a qualitatively different behaviour. As shown by Figure \ref{scatter}, these models are less accurate and the deviation of these models from the other curves in panels (d), (e) and (f) illustrates the limitation of these models.  
The fact that the significantly different models shown by the solid curve, the red dot-dashed curve and the orange dot-dashed curve agree in all panels (a) - (f) indicates that these models produce physical extrapolation.

\section{Conclusion}

GP regression is a powerful tool for building global PES of polyatomic systems. It is system agnostic and can be used to automate the construction of PES for applications in both classical and quantum molecular dynamics. It provides a global representation of the PES as well as the derivatives of the PES, which can be evaluated by differentiating the kernels in Eq. (\ref{GP-mean}). GP models also provide a Bayesian uncertainty of the PES, which can be used for Bayesian optimization in order to locate efficiently the extrema of the surface and evaluate the uncertainty of the dynamical observables determined by the PES \cite{BML}.  
The accuracy of GP models increases with the number of training points, as was shown in the context of PES, for example, in Ref. \cite{jie-jpb}. 
However, the numerical difficulty of training and evaluating GP models quickly increases with the number of energy points $n$ used for training. Because $n$ required for constructing accurate models increases with the number of degrees of freedom $N$,  
this makes the application of GP regression to polyatomic systems with a large number of degrees of freedom difficult. In particular, the numerical difficulty of training a GP model scales with the number of training points as ${\cal O}(n^3)$, which puts a limit on $n$ that can be used for practical applications. 
While training of an exact GP model with $n=3,000,000$ has been demonstrated \cite{GP-for-high-dimensions,GP-for-high-dimensions-2}, applications aiming at the construction of accurate PES are currently limited to $n \lesssim 10,000$. It is therefore important to find approaches that could be used to train accurate GP models with a small number of energy points. In the present work, we have demonstrated one such approach.  
   
In particular, we have shown that the accuracy of GP interpolation of PES with fixed $n$ can be enhanced by increasing the kernel complexity using a greedy search algorithm with the Bayesian information criterion as a model selection metric. 
%Our results show that models with a smaller number of training points $n$ benefit more from this accuracy improvement. 
%Thus, the error of the PES trained by $n=300$ drops by a factor of 2.6, while the error of the surface with $n=1000$ decreases by a factor of XXX, as the kernel complexity is increased. 
Using this approach, we have constructed a global, six-dimensional PES for H$_3$O$^+$ with the RMSE $< 100$ cm$^{-1}$ in the energy range from zero to 21,000 cm$^{-1}$ using only 500 energy points as inputs to the GP model. As the scaling of $n$ with $N$ is known to be linear, this suggests an approach that can be applied to systems with more degrees of freedom than currently feasible. In a separate work \cite{hiroki}, we have applied GP regression to building a global surface of a polyatomic molecule with 51 dimensions. 

We have also shown that the GP models with composite kernels can extrapolate the PES. We have demonstrated the extrapolation by training the GP models of PES by the {\it ab initio} points at low energy and testing the accuracy of the GP prediction of the PES at high energies. Our results indicate that the GP models produce a physical representation of the PES both within and outside the distributions of the training energy points. This makes GP models a valuable tool for exploring the landscape of unknown PES of complex molecular systems. For such systems, one can envision a procedure that beings with a GP model of the global surface based on a small number of {\it ab initio} points (e.g. 30 $\times$ the number of dimensions). The accuracy of the PES near desired PES features could then be enhanced by placing more energy points in the corresponding part of the configuration space. Because the number of energy points required to construct accurate representations of the PES is small, this approach can be used to obtain surfaces with very high-level {\it ab initio} calculations.

\section*{Acknowledgments}

We thank Rodrigo Vargas-Hernandez for useful discussions and technical assistance. This work is supported by NSERC of Canada.

\end{document}